\documentclass[preprint]{aastex}

\usepackage[dvips]{color}
\usepackage{rotating}
\usepackage{here}
\usepackage{amsmath,amssymb}
\usepackage{threeparttable}
\slugcomment{}
\shortauthors{Xue et al.}
\shorttitle{Parameter survey of coplanar flip}
\begin{document}
\title{Possible Outcomes of Coplanar High-eccentricity Migration: Hot Jupiters, Close-in Super-Earths, and Counter-orbiting Planets}
\author{
Yuxin Xue\altaffilmark{1},  Kento Masuda\altaffilmark{1,2,3}, Yasushi Suto\altaffilmark{1,4}
} 
\altaffiltext{1}{Department of Physics, The University of Tokyo, Tokyo
113-0033, Japan}

\altaffiltext{2}{Department of Astrophysical Sciences, Princeton University, Princeton, NJ 08544, USA}

\altaffiltext{3}{NASA Sagan Fellow}

\altaffiltext{4}{Research Center for the Early Universe, School of
Science, The University of Tokyo, Tokyo 113-0033, Japan}

\email{yuxin@utap.phys.s.u-tokyo.ac.jp}
\begin{abstract}

We investigate the formation of close-in planets in near-coplanar eccentric hierarchical triple systems via the secular interaction between an inner planet and an outer perturber (Coplanar High-eccentricity Migration; CHEM). We generalize the previous work on the analytical condition for successful CHEM for point masses interacting only through gravity by taking into account the finite mass effect of the inner planet. We find that efficient CHEM requires that the systems should have $m_1\ll m_0$ and $m_1\ll m_2$. In addition to the gravity for point masses, we examine the importance of the short-range forces, and provide an analytical estimate of the migration time scale. We perform a series of numerical simulations in CHEM for systems consisting of a sun-like central star, giant gas inner planet and planetary outer perturber, including the short-range forces and stellar and planetary dissipative tides. We find that most of such systems end up with a tidal disruption; a small fraction of the systems produce prograde hot Jupiters (HJs), but no retrograde one. In addition, we extend CHEM to super-Earth mass range, and show that the formation of close-in super-Earths in prograde orbits is also possible. Finally, we carry out CHEM simulation for the observed hierarchical triple and counter-orbiting HJ systems. We find that CHEM can explain a part of the former systems, but it is generally very difficult to reproduce counter-orbiting HJ systems.

\end{abstract}

\section{Introduction \label{sec:intro}}

Hundreds of hot Jupiters (hereafter, HJs) have been observed around main sequence stars. It is commonly believed that HJs formed at larger distances from their host stars (beyond the ice line) and subsequently migrated to the current location of less than 0.1 AU. The known migration scenarios include disk migration through the interaction with the protoplanetary disk \citep[e.g.,][]{Lin1996,Alibert2005} and high-$e$ migration, in which planets approach extremely high eccentricities and then suffer from tidal circularization. The possible mechanisms to form HJs in the latter scenario include (1) planet-planet scattering \citep[e.g.,][]{Rasio1996, Nagasawa2008, Nagasawa2011, Beauge2012}, (2) the Lidov-Kozai migration \citep[e.g.,][]{Lidov1962, Kozai1962, Wu2003, Fabrycky2007,Naoz2011,Naoz2012,Petrovich2015a, Anderson2016}, and (3) secular chaos \citep{Wu2011,Hamers2016}. Those different models may have contributed to the observed HJ population to some degree, but the dominant channel, if any, is still a matter of debate. 

The observed distribution of the orbital elements provides important clues on the dynamical origin of HJs and may distinguish among those models. Most of the observed HJs have low eccentricities and are preferentially located at $\sim0.04-0.05$ AU away from the central star. The projected spin-orbit angle, $\lambda$, the sky-projected angle between the spin of the central star and the orbit of the planet, has been measured for more than 90 transiting planets mainly through the Rossiter-McLaughlin (RM) effect. More than half of them are aligned, and about $40\%$ of them exhibit significant misalignment. Furthermore, there are 13 retrograde ($\lambda>90^{\circ}$) and 2 counter-orbiting ($\lambda>160^{\circ}$, just for definiteness in this paper)planets; HAT-P-6b with $\lambda = 165^\circ \pm {6^\circ}$ \citep{Albrecht2012} and HAT-P-14b with $\lambda = 189^{\circ}.1 \pm {5^\circ.1}$ \citep{Winn2011}\footnote{The projected spin-orbit angle, $\lambda$, differs from the {\it true} spin-orbit angle, $i_{s1}$, due to the projection effect. For $\lambda$ observed via RM effect, $\lambda<i_{s1}$, when $\lambda<90^{\circ}$; $\lambda>i_{s1}$, when $\lambda>90^{\circ}$. Thus, planetary systems with $\lambda>160^\circ$ may not be necessarily counter-orbiting, but just retrograde. Indeed, HAT-P-7b has $\lambda>160^{\circ}$, but turns out to be not counter-orbiting after the measurement of stellar inclination with asteroseismology \citep{Benomar2014}.}.

The above two candidates for counter-orbiting HJ systems are particularly interesting, since they cannot be explained by any of the above models. Disk migration predicts a low spin-orbit angle due to the quasi-Keplerian motion in a gaseous disk, but recent studies claim that disk migration may generate possible spin-orbit misalignment under certain situations \citep[e.g,][]{Bate2010,Spalding2014,Rogers2012}. In contrast, the other three high-$e$ migration mechanisms predict broadly distributed, even retrograde, spin-orbit angles. Even in those cases, however, counter-orbiting HJ systems are known to be difficult to form. Figure 14 of \citet{Nagasawa2011}, for instance, indicates that there is no counter-orbiting HJ in all the 241 HJs produced by planet-planet scattering. Figure 3 of \citet{Naoz2012} also illustrates that no counter-orbiting HJ is produced by the Lidov-Kozai migration for their ten thousand runs with a stellar perturber. Finally, the simulations by \citet{Hamers2016} show that all the HJs formed by secular chaos have the spin-orbit angle less than $120^{\circ}$ (see their Figure 10). 

Therefore, in this paper, we consider yet another possibility of high-$e$ migration mechanism that HJs form via the secular interaction between two orbits in a near-coplanar eccentric hierarchical triple configuration. Throughout this paper, we refer this HJ formation mechanism to {\it Coplanar High-eccentricity Migration (CHEM)} following \citet{Petrovich2015b}. CHEM is a potentially unique formation model of counter-orbiting HJs that no other model is known to generate. In reality, previous papers indicate that even CHEM is not so easy to produce counter-orbiting HJs. Nevertheless, even if this model is not able to produce count-orbiting HJs efficiently, it also provides a channel to produce prograde HJs. This is why we consider this model in this paper with particular emphasis on the application to the observed systems. 

As mentioned in the above, CHEM was originally proposed as a potential channel to form counter-orbiting HJs by \citet{Li2014}. They pointed out that the interaction due to the outer perturber can increase the eccentricity of the inner planet and flip its orbit by $\sim 180^{\circ}$. Such a counter-orbiting eccentric planet soon becomes circularized with subsequent tidal dissipation, and is expected to become a counter-orbiting HJ eventually. In particular, they analytically derived an extreme eccentricity condition that the eccentricity of the inner planet reaches unity, which results in its orbital flip. The condition is derived assuming the test particle limit ($m_1\ll m_0$, $m_1\ll m_2$), where $m_0$, $m_1$, and $m_2$ are the mass of the central star, inner planet, and outer perturber, respectively.

\citet{Petrovich2015b} examined CHEM in more details. First, he generalized the extreme eccentricity condition on the basis of the conservation of the potential energy in the planetary limit ($m_1\ll m_0$, $m_1\leq m_2$) instead of the test particle limit by \citet{Li2014}. In addition, he considered the eccentricity suppression due to general relativity (GR) and planetary non-dissipative tides. Then he found that those effects significantly limit the range of parameters that achieve the extreme eccentricity. Finally, he performed the numerical simulations of CHEM with a planetary outer perturber, and found that all the resulting HJs have low spin-orbit angles in prograde orbits. 

The initial conditions of his simulations, however, do not cover the relevant parameter space for the expected orbital flip. The fact motivated \citet{XueSuto2016} to perform a comprehensive parameter survey for CHEM with a sub-stellar outer perturber. They also found that most of the resulting HJs are prograde, but a very small fraction of them ends up with counter-orbiting. This is because their numerical simulations fully cover the relevant extreme eccentricity region in contrast to \citet{Petrovich2015b}. Nevertheless, they do not attempt to explain the numerical results from the analytical point of view, which is partly described in the present paper. 

In this paper, first we generalize the planetary limit by \citet{Petrovich2015b} taking into account the finite mass of the inner planet on the dynamics of the central star. Then, we present a more general form of the extreme eccentricity condition. We perform the numerical simulation as has been done by \citet{XueSuto2016} for a sub-stellar outer perturber; we systematically explore the fate of the inner planet in CHEM for a planetary outer perturber. Such a configuration is expected from dynamically unstable multi-planetary systems. Our initial condition homogeneously covers the relevant extreme eccentricity condition, and thus differs from \citet{Petrovich2015b}. We provide an analytical estimate of the migration time scale including the effect of short-range forces (GR, planetary non-dissipative tides, and planetary rotational distortion) that significantly change the dynamics of the orbital evolution \citep{Liu2015}. This estimate is useful in interpreting the present simulation results, and also in predicting the migration time scale for different situations analytically. Furthermore, while all previous studies of CHEM focus on the HJ formation, several super-Earths with semi-major axis less than 0.1 AU have been observed, and their origin still remains unknown. Therefore, we extend our simulation to an inner planet of super-Earth mass to see if CHEM can account for those very close-in super-Earths. Finally, we apply our simulations to the observed systems and examine to what extent CHEM can explain the existence of close-in planets in hierarchical triple and counter-orbiting HJ systems.

The rest of the paper is organized as follows. In section \ref{sec:ana}, we review the previous analytical results of the extreme eccentricity condition, and present our generalization of the extreme eccentricity condition. In section \ref{sec:pl}, we perform the numerical simulations for a giant gas inner planet with a planetary outer perturber including the short-range forces and dissipative tides. We also estimate the migration time scale analytically and compare it with the numerical results. In section \ref{sec:superearth}, we consider the case with an inner planet of super-Earth mass in CHEM. Section \ref{sec:pl} and \ref{sec:superearth} consider hypothetical systems for the systematic parameter survey. Instead, section \ref{sec:real} presents our application to the observed systems that may result from CHEM. Section \ref{sec:discussion} is devoted to summary and discussion of the present paper.

\section{Analytical Approach to Extreme Eccentricity Condition\label{sec:ana}}

\subsection{Extreme Eccentricity Condition in Previous Studies}\label{sec:pre-flip}

The extreme eccentricity condition is defined such that the eccentricity of the inner planet, $e_{1}$, reaches unity in a near-coplanar hierarchical triple system. Since the extreme eccentricity is associated with the $\sim 180^{\circ}$ orbital flip for point masses interacting only through gravity \citep{Li2014}, the above condition is often referred to as the flip condition, but we do not use the latter in order to avoid confusion. \citet{Petrovich2015b} derived the extreme eccentricity condition from the double time-averaged gravitational interaction potential. The potential expanded up to the octupole order $(a_1/a_2)^3$ can be written as \citep[e.g,][]{Petrovich2015a}: 
\begin{eqnarray} 
\label{eq:potential}
\phi_{} &=& \frac{\phi_{0}}{(1-e_{2}^{2})^{3/2}}[\frac{1}{2}(1-e_{1}^{2})
(\hat{\mathbf k}_{1}\cdot \hat{\mathbf k}_{2})^{2}+(e_{1}^{2}-\frac{1}{6})-\frac{5}{2}
({\mathbf e}_{1}\cdot \hat{\mathbf k}_{2})^{2}] \nonumber\\
&& + \frac{25\epsilon_{\rm oct}\phi_{0}}{16(1-e_{2}^{2})^{3/2}}\{ {\mathbf e}_{1}\cdot \hat{\mathbf e}_{2}
[(\frac{1}{5}-\frac{8}{5}e_{1}^{2})-(1-e_{1}^{2})(\hat{\mathbf k}_{1}\cdot \hat{\mathbf k}_{2})^{2}
+7({\mathbf e}_{1}\cdot \hat{\mathbf k}_{2})^{2}] \nonumber\\
&&-2(1-e_{1}^{2})(\hat{\mathbf k}_{1}\cdot \hat{\mathbf k}_{2})
({\mathbf e}_{1}\cdot \hat{\mathbf k}_{2})(\hat{\mathbf k}_{1}\cdot {\mathbf e}_{2}) \},\end{eqnarray}
where
\begin{eqnarray} 
 \phi_{0} &=& \frac{3G}{4}\frac{m_{0}m_{1}m_{2}}{m_{0}+m_{1}}\frac{a_{1}^{2}}{a_{2}^{3}},\\
\label{eq:eoct}
\epsilon_{\rm} &=&\frac{m_{0}-m_{1}}{m_{0}+m_{1}}\frac{a_{1}}{a_{2}}
\frac{e_{2}}{1-e_{2}^{2}}.
\end{eqnarray}
In the above expressions, we denote by $m$ the mass, $a$ the semi-major axis, $e$ the eccentricity, $\hat{\mathbf k}$ the unit orbital angular momentum vector, and $\mathbf e$ the Lenz vector. The subscripts 0, 1, and 2 refer to the central star, inner body, and outer perturber, respectively. Here we adopt the reference plane perpendicular to the total orbital angular momentum of the system. The first term of the right-hand-side in equation (\ref{eq:potential}) corresponds to the quadrupole term ($\propto(a_1/a_2)^2$) and the second term refers to the octupole contribution ($\propto(a_1/a_2)^3$). Thus $\epsilon_{\rm}$ defined by equation (\ref{eq:eoct}) gives the relative significance of the octupole term in the potential.

In the coplanar limit ($\hat{\mathbf k}_{i}\cdot \hat{\mathbf k}_{j}=1$, $\mathbf e_{i}\cdot \hat{\mathbf k}_{j}=0$, for $i,j=1,2$), equation (\ref{eq:potential}) reduces to
\begin{eqnarray} \label{eq:oct-potential}
\tilde{\phi}_{\rm}&\equiv& \frac{\phi_{}}{\phi_{0} } =\frac{e_{1}^{2}+2/3}{2(1-e_{2}^{2})^{3/2}} 
-\frac{5\epsilon_{\rm}}{16}\frac{4+3e_{1}^{2}}{(1-e_{2})^{3/2}}e_1\cos \varpi,
\end{eqnarray}
where $\varpi$ is defined as $\varpi \equiv \cos^{-1} \hat{\mathbf e}_{1} \cdot \hat{\mathbf e}_{2}$, the angle between the inner and outer unit Lenz vectors. In section \ref{sec:ana}, we assume the exact coplanarity of the system just for simplicity. The extreme eccentricity condition can be derived from the conservation of the potential energy up to the octupole order combined with the total orbital angular momentum conservation. 

In the test particle limit ($m_1\ll m_0$, $m_1\ll m_2$), $e_2$ is constant since the effect of $m_1$ on $m_2$ is neglected. Thus, the extreme eccentricity condition can be obtained by simply setting $e_{1,f}=1$ in equation (\ref{eq:oct-potential}). The result becomes
\begin{eqnarray} 
\label{eq:flipcondition} 
\epsilon_{\rm pl} > \frac{8}{5} 
\frac{1-e_{1,i}^{2}}{7\cos \varpi_{f}-e_{1,i}(4+3e_{1,i}^{2})\cos \varpi_{i}},
\end{eqnarray}
where
\begin{eqnarray} 
\label{eq:epsilondefinition} 
\epsilon_{\rm pl} \equiv \frac{a_{1}}{a_{2}}\frac{e_{2}}{1-e_{2}^{2}}.
\end{eqnarray}
Here $\epsilon_{\rm pl}$ is the reduced version of $\epsilon_{\rm}$ in equation (\ref{eq:eoct}) in the limit of $m_1\ll m_0$. The subscripts $i$ and $f$ refer to the initial and final states, respectively. Equation (\ref{eq:flipcondition}) was first derived by \citet{Li2014} using a slightly different approach. 

\citet{Petrovich2015b} generalized the extreme eccentricity condition in the planetary limit ($m_{1}\ll m_{0}$, $m_{1} \leq m_{2}$). In this limit, equation (\ref{eq:oct-potential}) reduces to equation (1) of \citet{Petrovich2015b}:
\begin{eqnarray} \label{eq:pl-limit}
\tilde{\phi}_{\rm pl} =\frac{e_{1}^{2}+2/3}{2(1-e_{2}^{2})^{3/2}} 
-\frac{5\epsilon_{\rm pl}}{16}\frac{4+3e_{1}^{2}}{(1-e_{2})^{3/2}}e_1\cos \varpi.
\end{eqnarray}
Assuming the conservation of the potential up to the octupole order, the extreme eccentricity condition where $e_{1,f}$ reaches unity is written as
\begin{eqnarray} 
\label{eq:petrovich11-pl} 
\tilde{\phi}_{\rm pl}(e_{1,i},e_{2,i},\epsilon_{\rm pl, i},\varpi_{i}=\pi)&=& 
\tilde{\phi}_{\rm pl}(e_{1,f}=1,e_{2,f},\epsilon_{\rm pl, f},\varpi_{f}),\end{eqnarray}
where we set $\varpi_{i}=\pi$ for definiteness. Equation (\ref{eq:petrovich11-pl}) can be numerically solved along with the angular momentum conservation. Then, we obtain the range of $\epsilon_{\rm pl,i}$ such that $e_{1,f}$ can reach unity as the function of $e_{1,i}$, $e_{2,i}$, and $\varpi_{f}$. The lower and upper boundaries of the condition correspond to $\varpi_{f} = 0$ and $\pi$ in practice. Thus, equation (\ref{eq:petrovich11-pl}) puts constraints on $\epsilon_{\rm pl, i}$ as:
\begin{eqnarray} 
\label{eq:epsilon-oct-pl} 
\epsilon_{\rm L} < \epsilon_{\rm pl,i} \equiv \frac{a_{1,i}}{a_{2,i}}\frac{e_{2,i}}{1-e_{2,i}^{2}}
<\epsilon_{\rm U}. 
\end{eqnarray}

\citet{Petrovich2015b} examined equation (\ref{eq:petrovich11-pl}) and found that the required $e_{2,i}$ for $e_{1,f}\to 1$ decreases with increasing $a_{1,i}/a_{2,i}$ and $e_{1,i}$. He also realized the existence of the upper boundary for the extreme eccentricity condition, corresponding to $\epsilon_{\rm U}$ in equation (\ref{eq:epsilon-oct-pl}). Nevertheless, he did not examine it in detail. In reality, however, $\epsilon_{\rm U}$ is very important to determine the extreme eccentricity region when $m_1$ is comparable to $m_2$ as we will present in the next subsection.

\subsection{Generalization of Extreme Eccentricity Condition \label{sec:flip}}

In this subsection, we generalize the results of \citet{Petrovich2015b} in two aspects. First, we consider the system in which the inner planet and the outer perturber have comparable masses. In that case, $\epsilon_{\rm U}$ becomes important, which is not carefully examined in \citet{Petrovich2015b}. Second, we take account of the dynamical effect of the inner body on the central star. 

We repeat the similar analysis as in subsection \ref{sec:pre-flip} following \citet{Petrovich2015b}. The potential in the coplanar limit is given by equation (\ref{eq:oct-potential}). Unlike in the planetary limit as considered in \citet{Petrovich2015b}, we retain the term $(m_{0}-m_{1})/(m_{0}+m_{1})$ to include the dynamical interaction of $m_1$ on $m_0$. Then, the constraints on $\epsilon_{\rm i}$ should be
\begin{eqnarray} 
\label{eq:epsilon-oct} 
\epsilon_{\rm L} < \epsilon_{\rm i} \equiv \left(\frac{m_{0}-m_{1}}{m_{0}+m_{1}}\right)\frac{a_{1,i}}{a_{2,i}}\frac{e_{2,i}}{1-e_{2,i}^{2}}
<\epsilon_{\rm U}.
\end{eqnarray}
The region of $\epsilon_i$ satisfying equation (\ref{eq:epsilon-oct}) is referred to as the extreme eccentricity region. 

\begin{figure}[t]
\begin{center}
\includegraphics[width=16cm]{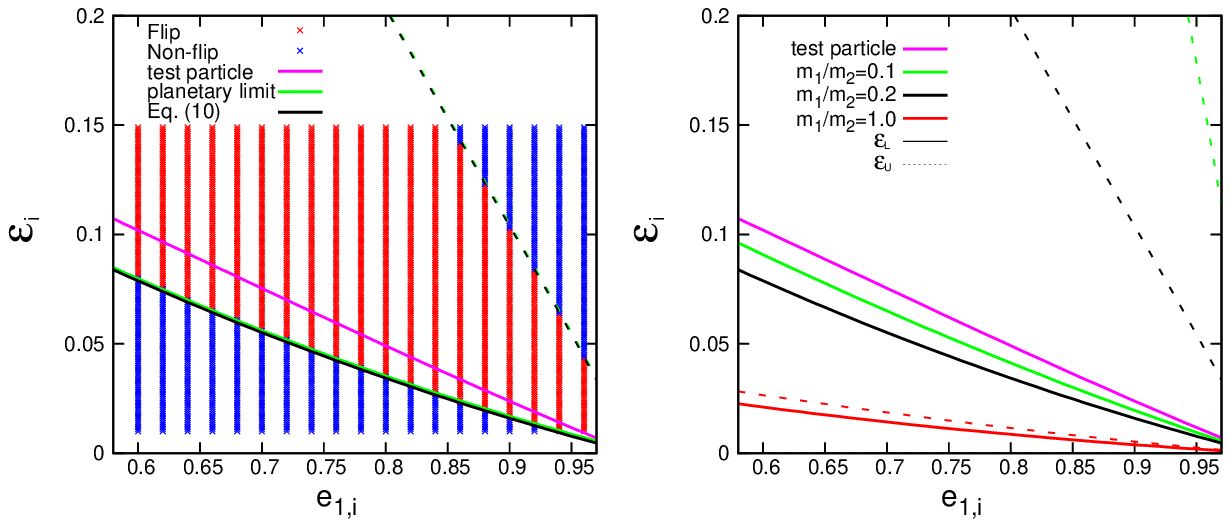} 
\caption{Left Panel: Comparison of the numerical results of the orbital evolution and the analytic extreme eccentricity conditions. The initial condition of the numerical simulation is $m_{0} = 1M_{\odot}$, $m_{1} = 1M_{\rm J}$, $m_{2}=5M_{\rm J}$, $a_2=50$ AU, $e_{2,i}=0.6$, $i_{12,i} = 6^{\circ}$, $\omega_{1,i} = 0$, $\omega_{2,i} = 0$, $\Omega_{1,i} = \pi$, $\Omega_{2,i} = 0$. The red crosses represent the flipped runs and the blue crosses are the non-flipped runs within $T_{\rm max}=10^{10}$ yrs. The black solid and dashed lines (Eq. (\ref{eq:epsilon-oct})) indicate the lower and upper boundaries of $\epsilon$ from equation (\ref{eq:epsilon-oct}), the green solid and dashed lines (planetary limit) plot the lower and upper boundaries of $\epsilon_{\rm pl}$ in the planetary limit taken from equation (\ref{eq:epsilon-oct-pl}), and the magenta line (test particle limit) is the extreme eccentricity condition in the test particle limit obtained from equation (\ref{eq:flipcondition}). Right panel: The extreme eccentricity condition from equation (\ref{eq:epsilon-oct}) for different values of $m_{2}$ with $m_0=1M_{\odot}$, $m_{1}=1M_{\rm J}$. The red, black, and green lines represent $m_2=1M_{\rm J}$ ($m_1/m_2=1.0$), $5M_{\rm J}$ ($m_1/m_2=0.2$), and $10M_{\rm J}$ ($m_1/m_2=0.1$); the solid and dashed lines refer to the lower and upper boundaries, respectively. The magenta line is the extreme eccentricity condition in the test particle limit, for which only the lower boundary exists.}\label{fig:flip1}
\end{center}
\end{figure}

In the left panel of Figure \ref{fig:flip1}, we compare the analytical and numerical results of the extreme eccentricity condition with $m_{0} = 1M_{\odot}$, $m_{1} = 1M_{\rm J}$, and $m_{2}=5M_{\rm J}$. The numerical simulation is basically identical to that described in section \ref{sec:initial}, but for point masses interacting only through gravity. The black, green, and magenta lines indicate the analytic boundaries derived from equation (\ref{eq:epsilon-oct}), in the planetary limit taken from equation (\ref{eq:epsilon-oct-pl}), and in the test particle limit derived from equation (\ref{eq:flipcondition}), respectively. The solid and dashed lines represent $\epsilon_{\rm L}$ and $\epsilon_{\rm U}$. In the simulation runs, we set the maximum simulation time $T_{\rm max}=10^{10}$ yr ($\sim 10^9$ orbital period) and stop each run before $T_{\rm max}$ if the system encounters the orbital flip. Here we use the orbital flip as the signal for the system reaching extreme eccentricity following \citet{Li2014}. The red crosses indicate the flipped runs and the blue crosses are the non-flipped runs. Our simulations homogeneously sample different $e_{1,i}$ and $a_{1,i}$ on ($e_{1,i}$, $\epsilon_i$) plane. We observe that the analytical criterion derived from equation (\ref{eq:epsilon-oct}) and in the planetary limit both are in good agreement with the numerical results, but the extreme eccentricity condition in the test particle limit is not sufficiently accurate; in the test particle limit, the predicted $\epsilon_{\rm L}$ is larger than the corresponding boundary in simulation and $\epsilon_{\rm U}$ does not exist. 

As shown in the left panel of Figure \ref{fig:flip1}, $\epsilon_{\rm U}$ significantly limits the extreme eccentricity region with $m_1=1M_{\rm J}$ and $m_{2}=5M_{\rm J}$. Then we consider the different choice of $m_2$ with particular emphasis on the importance of $\epsilon_{\rm U}$. The result is shown in the right panel of Figure \ref{fig:flip1} with solid and dashed lines representing $\epsilon_{\rm L}$ and $\epsilon_{\rm U}$. The red, black, and green lines are derived from equation (\ref{eq:epsilon-oct}) and correspond to  $m_2=1M_{\rm J}$, $5M_{\rm J}$, and $10M_{\rm J}$ with $m_0=1M_{\odot}$ and $m_1=1M_{\rm J}$, respectively. The magenta line is the extreme eccentricity condition in the test particle limit. The difference of the analytical estimate of equation (\ref{eq:epsilon-oct}) against that of equation (\ref{eq:flipcondition}) in the test particle limit increases as $m_1/m_2$ increases as expected. In particular, $\epsilon_{\rm U}$ does not show up in the test particle limit, but affects the extreme eccentricity region when taking account of the effect of $m_1$ on $m_2$. This modification becomes more important as $m_1/m_2$ increases; $\epsilon_{\rm U}$ only affects a small region for $m_2=10M_{\rm J}$ $(m_1/m_2=0.1)$, but significantly shrinks the extreme eccentricity region for $m_2=5M_{\rm J}$ $(m_1/m_2=0.2)$, and $1M_{\rm J}$ $(m_1/m_2=1.0)$. This implies that $\epsilon_{\rm U}$ can be safely neglected when $m_1 \ll m_2$, but should be carefully considered when $m_{1}$ becomes not negligible compared to $m_{2}$ ($m_1/m_2>0.1$). When the inner plant reaches the extreme eccentricity, the back reaction on $e_2$ increases with both $m_1/m_2$ and $a_1$ due to the orbital angular momentum conservation. Thus, both $\epsilon_{\rm L}$ and $\epsilon_{\rm U}$ shift towards smaller $a_1$ regions as $m_1/m_2$ increases.  

\begin{figure}[t]
\begin{center}
\includegraphics[width=18cm]{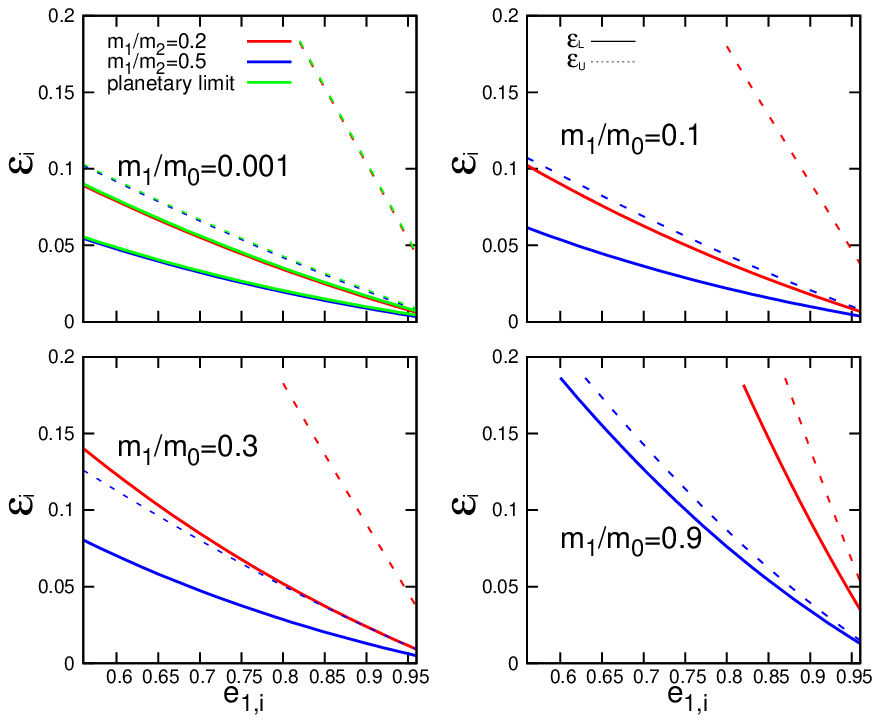} 
\caption{Extreme eccentricity condition for the different ratio of $m_{1}/m_0$ taken from equation (\ref{eq:epsilon-oct}). Four different cases with $m_1/m_0=0.001$, 0.1, 0.3, and 0.9 are presented, from left to right, upper to bottom. The red and blue lines correspond to $m_1/m_2=0.2$, and $0.5$; the solid and dashed lines refer to the lower and upper boundaries of $\epsilon$, respectively. The green lines in the left upper panel represent the boundaries of $\epsilon_{\rm pl}$ in the planetary limit.} \label{fig:flip2}
\end{center}
\end{figure}

Next, we consider the dependence of the extreme eccentricity condition derived from equation (\ref{eq:epsilon-oct}) on $m_1/m_0$, which is illustrated in Figure \ref{fig:flip2}. We present four different cases with $m_1/m_0=0.001,$ 0.1, 0.3, and $0.9$, so as to basically cover the range of a gas giant, sub-stellar object, M-dwarf and sun-like star with a sun-like central star, respectively. The lower and upper boundaries of $\epsilon$ are indicated by the solid and dashed lines. The red and blue lines correspond to $m_1/m_2=0.2$ and 0.5, respectively. The green lines present the boundaries in the planetary limit. In the left upper panel with $m_1/m_0=0.001$, the resulting extreme eccentricity region derived from equation (\ref{eq:epsilon-oct}) and from equation (\ref{eq:epsilon-oct-pl}) in the planetary limit are almost identical. As $m_1/m_0$ increases, the planetary limit becomes less accurate. As shown in the other three panels, for $m_1/m_0\geq0.1$, the extreme eccentricity region tends to be significantly narrower and shifted towards higher $e_{1,i}$ regime. Since the relative importance of the octupole term is proportional to $(m_{0}-m_{1})/(m_{0}+m_{1})$, increasing of $m_1/m_0$ corresponds to decreasing of the octupole effect. Therefore, in order to reach the extreme eccentricity, the inner orbit requires a higher $e_{1,i}$ to compensate the relatively smaller octupole term. 

In summary, $\epsilon_{\rm U}$ limits the extreme eccentricity region when $m_1$ becomes comparable with $m_2$. In section \ref{sec:pl}, we consider the systems with a giant gas inner planets and a planetary outer perturber, carefully considering the effect of $\epsilon_{\rm U}$. The planetary limit is a reasonably good approximation for $m_1\ll m_0$, but is not sufficiently accurate when $m_1/m_0\geq0.1$. Indeed, increasing $m_1/m_0$ and $m_1/m_2$ both shrink the extreme eccentricity region. Therefore, it is very difficult to form close-in orbit in CHEM for systems with a relatively massive inner body, for example, triple star systems.

\subsection{Short-range Force Effect\label{sec:srf-petrovich}}

So far, we have presented the extreme eccentricity condition neglecting the short-range forces. In reality, general relativity (GR), stellar and planetary non-dissipative tides, and stellar and planetary rotational distortion should be included as the short-range forces. Indeed, those short-range forces may significantly affect the dynamical evolution of the inner planet. In general, the short-range forces induce the additional precession of the pericenter of the inner planet, which suppresses the extreme eccentricity growth. Adopting the methodology similar to \citet{Petrovich2015b}, we analytically consider the short-range force effect including GR, planetary non-dissipative tides and planetary rotational distortion\footnote{We neglect the central stellar tides and rotational distortion, because their effects are indeed negligible in practice.}. In this case, the total dimensionless potential is given by 
\begin{eqnarray} 
\label{eq:potential-srf}
\tilde{\phi}_{\rm total}=\tilde{\phi}_{\rm}+\tilde{\phi}_{\rm GR}+\tilde{\phi}_{\rm TD}+\tilde{\phi}_{\rm PRD}.
\end{eqnarray}
In equation (\ref{eq:potential-srf}), $\tilde{\phi}_{\rm GR}$, $\tilde{\phi}_{\rm TD}$, and $\tilde{\phi}_{\rm PRD}$ refer to the dimensionless potential due to GR, planetary non-dissipative tides, and planetary rotational distortion: 
\begin{eqnarray} 
\label{eq:GR-pre}
 \tilde{\phi}_{\rm GR}&=&\frac{4Gm_0^2a_2^3}{c^2a_1^4m_2}\frac{1}{(1-e_1^2)^{1/2}},\\
\label{eq:TD-pre} 
\tilde{\phi}_{\rm TD}&=&\frac{4k_{2,1}}{3}\left(\frac{m_0^2m_2}{m_1}\right)\left(\frac{a_2^3R_1^5}{a_1^8}\right)\frac{1+3e_1^2+3e_1^4/8}{(1-e_1^2)^{9/2}},\\
\label{eq:PRD-pre} 
\tilde{\phi}_{\rm PRD}&=&\frac{2k_{2,1}}{9G}\left(\frac{m_0}{m_1m_2}\right)\left(\frac{a_2^3R_1^5}{a_1^5}\right)\frac{\omega_{\rm p}^2}{(1-e_1^2)^{3/2}},
\end{eqnarray}
where $c$ is the speed of light, $k_{2,1}$ is the second Love number of the inner planet, $R_1$ is the radius of the inner planet, and $\omega_{\rm p}$ is the spin rate of the inner planet. The potential energy we consider differs from \citet{Petrovich2015b} by including the effect of planetary rotational distortion, equation (\ref{eq:PRD-pre}). Although $\tilde{\phi}_{\rm PRD}$ turns out to make a minor contribution \citep[see Figure 5 in][]{XueSuto2016}, we include it for completeness. 

We repeat the similar analysis as in subsection \ref{sec:pre-flip} including those three short-range forces. The eccentricity of the inner planet, $e_{1}$, cannot reach exactly unity due to the short-range force effect (equations (\ref{eq:GR-pre}) to (\ref{eq:PRD-pre})). Instead, the maximum eccentricity of the inner planet, $e_{1,f}$, can be determined by the conservation of the potential as follows,
\begin{eqnarray} 
\label{eq:petrovich19} 
\tilde{\phi}_{\rm total}(e_{1,i},e_{2,i},\epsilon_i,\varpi_{i}=\pi)&=& 
\tilde{\phi}_{\rm total}(e_{1,f},e_{2,f},\epsilon_f,\varpi_{f}).\end{eqnarray}
The corresponding minimum pericenter distance of the inner planet, $q_{1,\rm min}$, can be inferred from $e_{1,f}$ as 
\begin{eqnarray} 
\label{eq:qmin} 
q_{1,\rm min} = a_{1}(1-e_{1,f}).\end{eqnarray}
Here $q_{1,\rm min}$ is related to the fate of the system, since tides are very sensitive to the pericenter distance. In order to produce a HJ, the inner planet must reach a sufficiently small pericenter distance, such that tidal dissipation can reduce the planet's semi-major axis within a few Gyrs, and its pericenter distance should be larger than the Roche limit to avoid disruption. 

We compute the lower and upper boundaries of $\epsilon_i$ that correspond to a given value of $q_{1,\rm min}$ as follows. We substitute the value of $e_{1,f}$  corresponding to $q_{1,\rm min}$ into equation (\ref{eq:petrovich19}). Combining with the orbital angular momentum conservation, we obtain $\epsilon_{\rm L}$ for $\varpi_f=0$, and $\epsilon_{\rm U}$ for $\varpi_f=\pi$ by solving equation (\ref{eq:petrovich19}) for $\epsilon_i$. The example of $\epsilon_{\rm L}$ and $\epsilon_{\rm U}$ for $q_{1,\rm min}$ are plotted in Figure \ref{fig:plfid} below.

As demonstrated by previous studies, the short-range force effect significantly affects the orbital evolution. Thus, it is necessary to include the short-range force effect to analytically interpret the numerical results. The detailed consideration will be presented in subsection \ref{sec:ana-fid}.

\section{Giant Gas Inner Planet with a Planetary Outer Perturber\label{sec:pl}}

So far, we have presented the extreme eccentricity condition analytically. In this section, we perform a series of numerical simulations to study the orbital evolution of hierarchical triple systems in CHEM with a giant gas inner planet and a planetary outer perturber, similarly as our previous paper with a sub-stellar outer perturber \citep{XueSuto2016}. We first present the model parameters and our fiducial case in subsection \ref{sec:initial}, and then consider how to interpret the numerical results in terms of the analytical argument in subsection \ref{sec:ana-fid}. Finally we discuss the parameter dependence and the final distribution of the orbital elements in subsections \ref{sec:sim-dep} and \ref{sec:af-dis}. 

\subsection{Simulation Parameters and Fiducial Results\label{sec:initial}}

We consider the evolution of near-coplanar hierarchical triple systems consisting of a sun-like central star ($m_0=1M_{\odot}$ and $R_0=1R_{\odot}$), a giant gas inner planet ($m_1=1M_{\rm J}$ and $R_1=1R_{\rm J}$), and a distant outer perturber ($m_2$). In our previous paper, we considered the sub-stellar perturber ($10M_{\rm J}<m_2<1M_{\odot}$), but here we consider the planetary perturber ($1M_{\rm J}<m_2<10M_{\rm J}$). In this subsection, we consider $m_2=5M_{\rm J}$ as the fiducial value and discuss the $m_2$ dependence in subsection \ref{sec:sim-dep}. This fiducial choice of masses aims to cover the relevant extreme eccentricity condition for CHEM to work (see the right panel of Figure \ref{fig:flip1}). It is in contrast with \citet{Petrovich2015b} who adopted $m_1=1M_{\rm J}$ and $m_2\in[1.3,1.7]M_{\rm J}$. Because of the upper boundary $\epsilon_{\rm U}$, his choice significantly limits the region of CHEM. This is mainly why more than 96\% of his simulation runs end up with non-migrating planets (see his Figure 7). In addition to the gravitational interaction up to the octupole order, our simulations include the short-range forces, and stellar and planetary dissipative tides. To be specific, the short-range forces incorporated are GR, stellar and planetary non-dissipative tides, and stellar and planetary rotational distortion. 

We solve the equations of motion described by \citet{Correia2011} and \citet{Liu2015}; see also Appendix A of \citet{XueSuto2016}. Our simulation models are basically specified by seven parameters, $m_2$, $a_{2,i}$, $e_{2,i}$, $i_{12,i}$, $t_{\rm v,p}$, $f$, and $m_0$. Here $i_{12}$ is the mutual orbital inclination, $t_{\rm v,p}$ is the viscous time scale\footnote{The control parameter of equilibrium tide model can be specified by the quality factor $Q$ or the tidal delay time $\Delta t$ instead of the viscous time scale $t_{v}$. $\Delta t$ is related to $t_{v}$ by $\Delta t=3(1+k_{2})R_{}^3/(Gm_{}t_{v})$, and $Q$ is related to $\Delta t$ by $1/Q = 2|\omega-n|\Delta t$, where $k_{2}$ is the second Love number, $\omega$ is the spin rate, and $n$ is mean motion.} of the inner planet, and $f$ is the disruption control parameter in terms of the Roche limit, $R_{\rm roche} \equiv f(m_{0}/m_{1})^{1/3}R_1$. In the current simulation, we adopt a tidal model in which $\Delta t_{\rm p}$ $(t_{\rm v,p})$ is constant. In this case, the tidal quality factor, $Q_{\rm p}$, is time-dependent as
\begin{eqnarray} 
\label{eq:Q-tide} 
Q_{\rm p}&=&\frac{1}{2|\omega_{\rm p}-n_1|\Delta t}\\\nonumber
&=&\frac{Gm_{1}t_{\rm v,p}}{6(1+k_{2,1})R_{1}^3|\omega_{\rm p}-n_1|}\\ \nonumber
&\approx& 3\times10^3\left(\frac{t_{\rm v,p}}{0.03 \rm yr}\right)\left(\frac{P_{\rm p}}{0.5 \rm day}\right)\left(\frac{m_1}{M_{\rm J}}\right)\left(\frac{R_{\rm J}}{R_1}\right)^3,
\end{eqnarray}
where $P_{\rm p}$ is the orbital period of the inner planet. Because the spin rate of the inner planet, $\omega_{\rm p}$, soon arrives at its equilibrium state, $Q_{\rm p}$ indeed becomes approximately one magnitude larger in a very short time scale. We fix the viscous time scale of the central star to 50 yr, and the second Love number of the central star and inner planet to 0.028 and 0.5, respectively \citep[e.g.,][]{Correia2011}. The initial spin period for the central star and inner planet are set to 25 days and 0.5 days based on our Sun and Jupiter. The initial phase angles for the argument of pericenter $\omega$ and the longitude of ascending node $\Omega$ are fixed to $\omega_{1} = 0^{\circ}$, $\omega_{2} = 0^{\circ}$, $\Omega_{1} = 180^{\circ}$, $\Omega_{2} = 0^{\circ}$ for definiteness. Since planets are generally expected to form in a protoplanetary disk nearly perpendicular to the stellar spin axis, the spin-orbit angle, $i_{s1}$, is initially set to 0. We do not assume any prior distribution of the orbital elements for $a_{2,i},e_{2,i}$, and $i_{12,i}$ because of the difficulty to estimate the probability for actual near-coplanar hierarchical two-planet systems. The parameter dependence will be discussed in subsection \ref{sec:sim-dep}.

We run $\sim 2000$ simulations over the grids of ($e_{1,i}$, $\epsilon_i$) for our fiducial models; $e_{1,i}$ is varied between $0.6$ and $0.96$ with a constant interval of $0.02$, and  $\epsilon_i$ is varied between $\epsilon_{\rm L}$ derived from equation (\ref{eq:epsilon-oct}) and $0.15$ with a constant interval of $0.001$. The upper limit of 0.15 guarantees the validity of the secular approximation. We are interested in the extreme eccentricity region where the inner planet suffers from strong orbital evolution. Thus, the region with $\epsilon_i<\epsilon_{\rm L}$ is not simulated due to their regular orbital evolution ({\it {i.e.}}, no migration). 

We perform each simulation up to the maximum simulation time, $T_{\rm max}=10^{10}$ yr. Even before $T_{\rm max}$, we stop the simulation when the inner planet satisfies both $a_{1,f} < 0.1$ AU and $e_{1,f} < 0.01$, which we regard to be a HJ. We also stop the simulation when the pericenter distance of the inner planet, $q_1=a_1(1-e_1)$, reaches less than the Roche limit. The time at which the simulation is stopped is referred to as the stopping time, $T_{\rm s}$. Following the definition of \citet{XueSuto2016}, the outcomes of our simulations are divided into four categories as shown in Table \ref{tab:outcome1}. The model parameters and fraction of final fates are summarized in Table \ref{tab:pl}. We first discuss the fiducial case and its implications, then move to the parameter dependence and resulting distribution of the orbital elements. 

\begin{table}[t]
\begin{center}
\caption{Four categories of the final outcomes for systems described in section \ref{sec:pl}.}
\begin{threeparttable}
\begin{tabular}{|l||l|} 
\hline
Category & Condition \\ \hline 
\textbf{PHJ} (prograde HJ) &  $a_{1,f} < 0.1$ AU, $e_{1,f} < 0.01$ and $i _{12,f}<\pi/2$.\\ \hline
\textbf{RHJ} (retrograde HJ) &  $a_{1,f} < 0.1$ AU, $e_{1,f} < 0.01$ and $i _{12,f}>\pi/2$. \\ \hline
\textbf{TD} (tidally disrupted & The pericenter distance of the inner planet \\ 
planet)&$q_1\equiv a_1(1-e_1)$ reaches less than the Roche limit:\\
&$q_1 < R_{\rm roche} \equiv f(m_{0}/m_{1})^{1/3}R_1
\approx 0.0126 \left(\frac{f}{2.7}\right){\rm AU}.$\\
&Our fiducial value of  $f$ is 2.7 \citep[e.g,][]{Guillochon2011}, \\
& and we also consider $f=2.16$ \citep{Faber2005} for another \\
& possibility due to the uncertainty of this appropriate value.\\ \hline
\textbf{NM} (non-migrating & The inner planet does not exhibit a significant migration,\\ 
planet) & and stays at an orbit with $a_{1,f} \sim a_{1,i}$ until $t=T_{\rm max}$.\\ \hline
\end{tabular}
\end{threeparttable}
\label{tab:outcome1}
\end{center}
\end{table}

\begin{table}[t]
\begin{center}
\caption{Summary of parameters and the fates of our simulation runs described in section \ref{sec:pl}.}
\begin{threeparttable}
\begin{tabular}{|c||c|c|c|c|c||c|c|c|c|} 
\hline
Model &$a_{2}$ & $m_{2}$&$i_{12}$& $t_{\rm v, \rm p}$ &f&PHJ&RHJ&NM&TD\\ 
 &au&$M_{\rm J}$&&yr&&&&& \\  \hline 
fiducial&$50$& 5&$6^{\circ}$ & 0.03&2.7&24.7\%&0.0\%&13.6\%&61.7\%\\ \hline
a200&$200$& 5&$6^{\circ}$ & 0.03&2.7&27.5\%&0.5\%&19.4\%&52.7\%\\\hline

m10&$50$& 10 &$6^{\circ}$ & 0.03&2.7&19.4\%&0.0\%&3.9\%&76.7\%\\
m07&$50$& 7&$6^{\circ}$ & 0.03&2.7&22.5\%&0.0\%&4.3\%&73.3\%\\
m06&$50$& 6 &$6^{\circ}$ & 0.03&2.7&23.8\%&0.0\%&9.1\%&67.1\%\\
m04&$50$&4 &$6^{\circ}$ & 0.03&2.7&29.9\%&0.0\%&19.0\%&51.2\%\\
m03&$50$& 3&$6^{\circ}$ & 0.03&2.7&35.3\%&0.0\%&40.1\%&24.6\%\\
m02&$50$&2 &$6^{\circ}$ & 0.03&2.7&37.0\%&0.0\%&62.9\%&0.1\%\\
m01&$50$& 1 &$6^{\circ}$ & 0.03&2.7&5.7\%&0.0\%&94.2\%&0.1\%\\ \hline

i30&$50$& 5&$30^{\circ}$ & 0.03&2.7&13.6\%&0.0\%&11.7\%&74.8\%\\\hline
t003&$50$& 5 &$6^{\circ}$ & 0.003&2.7&60.1\%&0.0\%&11.3\%&28.6\%\\ \hline
f216&50&5 &$6^{\circ}$ & 0.03&2.16&86.3\%&0.0\%&13.6\%&0.1\%\\\hline
\end{tabular}
\begin{tablenotes}[para,flushleft]
\item{\textbf{Note.} All the models adopt $m_0=1M_{\odot}$, $m_{1} = 1M_{\rm J}$, $r_{1} = 1R_{\rm J}$,  $\omega_{1,i} = 0^{\circ}$, $\omega_{2,i} = 0^{\circ}$, $\Omega_{1,i} = 180^{\circ}$, $\Omega_{2,i} = 0^{\circ}$, and $i_{s1,i} = 0^{\circ}$. The second Love number, $k_2$, of the central star and inner planet are set to 0.028 and 0.5, respectively. The outcomes are divided into four categories: prograde HJ (PHJ; $a<0.1$ AU, $e<0.01$, and $i_{\rm 12}<90^{\circ}$), retrograde HJ (RHJ; $a<0.1$ AU, $e<0.01$, and $i_{\rm 12}>90^{\circ}$), non-migrating planet (NM; $T_{\rm s}=10^{10}$ yr) and tidally disrupted planet (TD; $q_1<R_{\rm roche}$). For each model, we perform $\sim2000$ runs by varying ($e_{1,i}$, $\epsilon_{i}$) systematically.} 
\end{tablenotes}
\end{threeparttable}
\label{tab:pl}
\end{center}
\end{table}

\begin{figure}[t]
\begin{center}
\includegraphics[width=16cm]{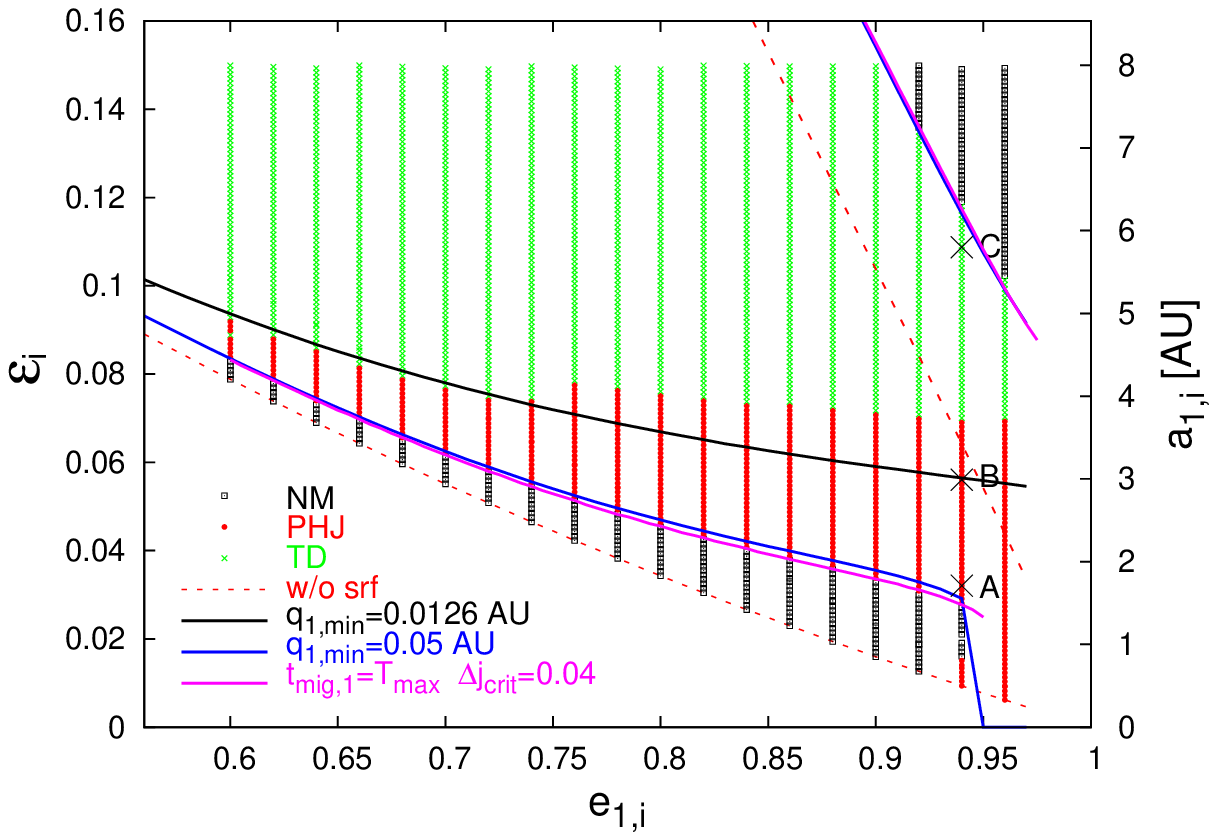} 
\caption{Results of the inner planet on ($e_{1,i}, \epsilon_{1,i}$) plane for the fiducial case in systems with a giant gas inner planet and a planetary outer perturber: $m_0=1M_{\odot}, m_{1}=1M_{\rm J}, m_{2}=5M_{\rm J}, e_{2,i}=0.6, a_2=50$ AU, $t_{\rm v,p} = 0.03$ yr, and $f=2.7$. $e_{1,i}$ is varied between $0.6$ and $0.96$ with a constant interval of $0.02$, and $\epsilon_i$ is varied between $\epsilon_{\rm L}$ derived from equation (\ref{eq:epsilon-oct}) and $0.15$ with a constant interval of $0.001$. The final states are indicated by black open squares for non-migrating planets (NM), red filled circles for prograde HJs (PHJ), and green crosses for tidally disrupted planets (TD), respectively. The red dashed lines correspond to the extreme eccentricity condition derived from equation (\ref{eq:epsilon-oct}). The blue and black lines correspond to the desired $\epsilon_i$ ($a_{1,i}$) to reach $q_{1,\rm min} = 0.05$ and $0.0126$ AU including the short-range force effect. The magenta lines correspond to the migration boundary ($t_{\rm mig,1}=10^{10}$ yrs) of the analytical estimate derived from equations (\ref{eq:t-all}), (\ref{eq:t-eo}), and (\ref{eq:t-cir}) with $\Delta j_{\rm crit} = 0.04$. The points A, B, and C correspond to Figure \ref{fig:ple94prosample1}a, b, and c, respectively.} \label{fig:plfid}
\end{center}
\end{figure}

Figure \ref{fig:plfid} shows the numerical results and analytical predictions of our fiducial case on $(e_{1,i}, \epsilon_i)$ plane with $m_0=1M_{\odot}$, $a_2=50$ AU, $m_2=5M_{\rm J}$, $e_{2,i}=0.6$, $i_{12,i}=6^{\circ}$, $t_{\rm v,p} = 0.03$ yr, and $f=2.7$. The lower and upper boundaries of $\epsilon$ analytically estimated from equation (\ref{eq:epsilon-oct}) are plotted in red dashed lines. Those analytical estimates are different from our numerical results, because the former neglects the short-range force effect. In this figure, we show non-migrating planets (NM), prograde HJs (PHJ), and tidally disrupted planets (TD) in black open squares, red filled circles, and green crosses, respectively. 

The most common outcome is tidally disrupted planets (TD; 61.7\%). This outcome is preferentially located in the region in which the inner planet has relatively large $\epsilon_i$ and therefore $a_{1,i}$, since the stronger gravitational interaction due to the outer perturber produces a more extreme eccentricity. 

The second common outcome is prograde HJs (PHJ; 24.7\%). PHJs are located in the region between NM and TD. We emphasize that the current simulation produces no RHJ unlike the case with a sub-stellar perturber \citep{XueSuto2016}. First, we note that the strength of tides is determined by the pericenter distance of the inner planet, $q_1=a_1(1-e_1)$. If $a_1$ is smaller, $e_1$ does not have to be so close to unity for the efficient tidal circularization. On the other hand, the orbital flip requires very extreme eccentricity. In the current simulations, we consider a planetary perturber, and therefore we adopt smaller $a_{1,i}$ relative to the previous study to a sub-stellar perturber. Thus, the tidal circularization always happens before the orbital flip. As a result, no RHJ is produced. In the fiducial case, we adopt the disruption radius $\approx0.0126$ AU with $f=2.7$, thus one may expect the $q_{1,\rm min}=0.0126$ AU line as the boundary between PHJ and TD. The black line corresponds to our analytical estimate of $q_{1,\rm min}=0.0126$ AU derived from equation (\ref{eq:petrovich19}), which is qualitatively consistent with the simulation results. The detailed consideration is presented in subsection \ref{sec:bound-PHJTD}.

Non-migrating planets (NM; 13.6\%) are located around the edge of bottom-left and upper-right region in Figure \ref{fig:plfid}, but the migration boundaries between NM and PHJ/TD do not simply follow the analytical extreme eccentricity condition (red dashed lines). Instead, the migration boundaries between NM and PHJ/TD are determined by the maximum simulation time that we adopted ({\it {i.e.}}, $T_{\rm max}=10^{10}$ yr). The detailed analytical consideration is presented in subsection \ref{sec:mig-time}.

In summary, our simulations indicate that CHEM cannot produce counter-orbiting HJs, but produces PHJs to some degree for systems with a giant gas inner planet and a planetary outer perturber. We also reconfirm that the short-range forces are important on the orbital evolution of the inner planet in such systems. The detailed analytical consideration incorporating the short-range forces will be presented in the next subsection.

\subsection{Analytical Interpretation of the Fiducial Case\label{sec:ana-fid}}
 
\subsubsection{Boundary between PHJ and TD\label{sec:bound-PHJTD}}

The boundary between PHJ and TD is determined basically by the Roche limit. In the fiducial case, we adopt $f=2.7$ and thus the Roche limit becomes $\approx0.0126$ AU. We attempt to set $q_{1,\rm min}=R_{\rm roche}$ derived from equation (\ref{eq:petrovich19}) with $\varpi_f=0$ to analytically estimate this boundary on ($e_{1,i}, \epsilon_{1,i}$) plane. The result is plotted with the black solid line in Figure \ref{fig:plfid}.

We observe that our analytical estimate of this boundary is qualitatively in agreement with the numerical results. This analytical estimate is accurate for $e_{1,i}<0.75$, but becomes less accurate for $e_{1,i}>0.75$; in the latter case, our simulations indicate that HJs can form in the region where $q_{1,\rm min}<R_{\rm roche}$. 

The following two effects change the analytical estimate of the boundary of $q_{1,\rm min}$ on the basis of equation (\ref{eq:petrovich19}), which may account for the discrepancy. One is dissipative tides, which decreases $e_1$ further. Thus, the actual $q_{1,\rm min}$ tends to be larger than the analytical estimate based on equation (\ref{eq:petrovich19}), and the boundary between PHJ and TD moves upwards. The importance of dissipative tides increases for a higher $e_{1,i}$, since the inner planet spends more time in a high-$e$ phase. The other is the break down of the exact coplanar condition. Equation (\ref{eq:petrovich19}) is based on the assumption of the exact coplanar orbit, but we adopt $i_{12,i}=6^{\circ}$ in the fiducial case of our simulations. The non-zero $i_{12,i}$ increases the maximum achievable eccentricity due to the angular momentum conservation, and moves the boundary between PHJ and TD downwards. Since the amplitude of the inclination oscillation increases as $e_{1,i}$ decreases, this effect plays a more important role in the lower $e_{1,i}$ region.

\subsubsection{Boundaries between NM and PHJ/TD\label{sec:mig-time}}

\begin{figure}[t]
\begin{center}
\includegraphics[width=16cm]{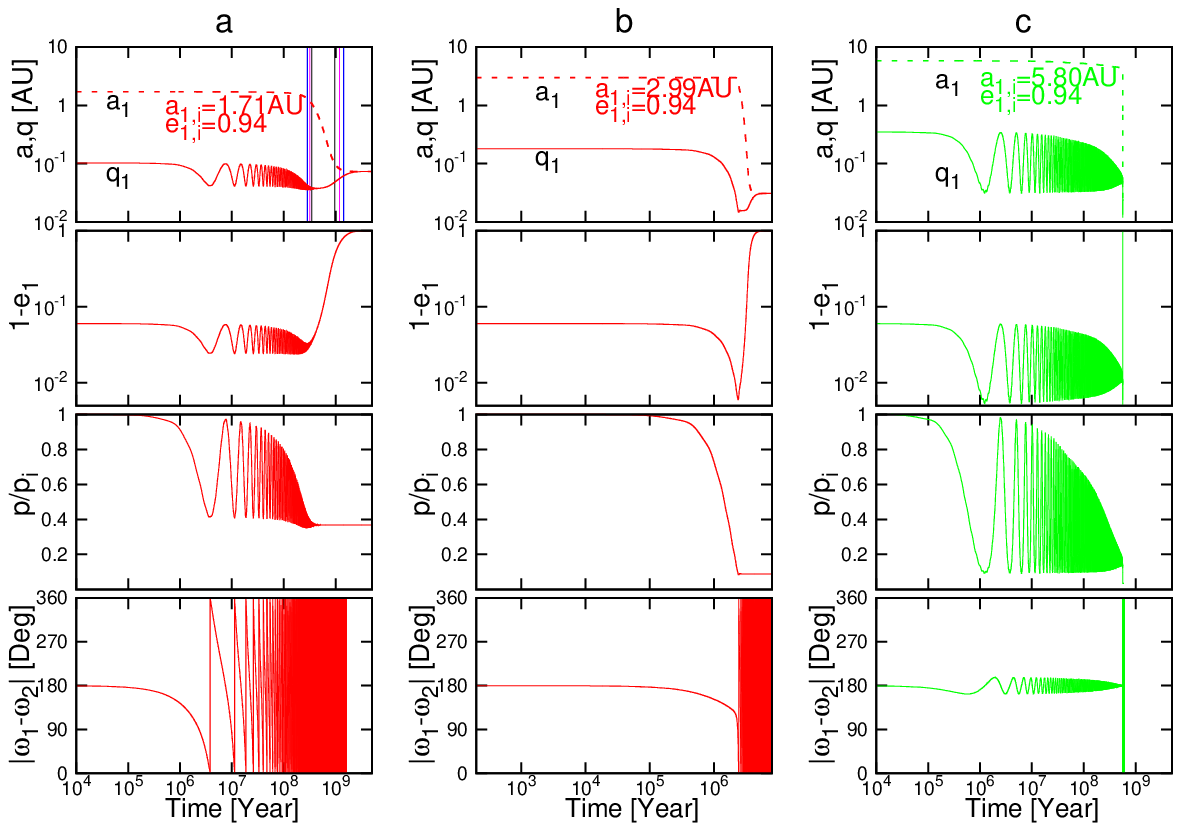} 
\caption{Orbital evolution of PHJ and TD in the fiducial case with $e_{1,i} = 0.94$ for three different $a_{1,i}$. The evolution of $a_1$ and $q_1$, $e_1$, $p/p_{i}$, and $|\omega_1-\omega_2|$ are presented from top to bottom, where $a_{1}$ is plotted in dashed lines, $q_1$, $e_{1}$, $p/p_{i}$, and $|\omega_1-\omega_2|$ are shown in solid lines, respectively. Panel a is $a_{1,i}=1.71$ AU which corresponds to the region close to the lower migration boundary (point A in Figure \ref{fig:plfid}); Panel b has $a_{1,i}=2.99$ AU, which is inside the PHJ region with intermediate $\epsilon_i$ (point B in Figure \ref{fig:plfid}); Panel c is $a_{1,i}=5.80$ AU, which corresponds to the tidal disruption region near the upper migration boundary (point C in Figure \ref{fig:plfid}). The solid vertical lines in the upper plot of Panel a refer to our analytical estimate of the migration boundary of $t_{\rm eo}$ and $t_{\rm mig,1}$, where black, magenta, and blue lines correspond to $\Delta j_{\rm crit}=0.03, 0.04$, and 0.05, respectively.} \label{fig:ple94prosample1}
\end{center}
\end{figure}

The lower boundary between NM and PHJ is determined by the epoch when the inner planet migrates to become a HJ within $T_{\rm max}=10^{10}$ yr. The upper boundary between NM and TD is similarly determined, but in this case, the orbital angular momentum loss of the inner planet is rapid, and thus there is no stable HJ in between. We will explain the behavior in detail in Figure \ref{fig:ple94prosample1} later. As shown in Figure \ref{fig:plfid}, $q_{1,\rm min}=0.05$ AU plotted in blue solid lines empirically shows good agreement with both the lower and upper migration boundaries in the numerical simulation. This consistency may reflect some correlation between $q_{1,\rm min}$ and the migration time scale, the time scale for the planet becoming a HJ. 

Indeed, the basic mechanism to determine the migration boundary is tides, which are very sensitive to $q_{1,\rm min}$. A more precise approach to determine the migration boundary is to use the equations of motion directly. The migration time scale is determined by the evolution equations of $a_1$ and $e_1$ governed by planetary tides, which are given as \citep{Correia2011}
\begin{eqnarray} 
\label{eq:adot} 
\dot{a}_1&=&2\frac{\tilde{K_1}}{a_{1}^{7}}\left(f_{2}(e_{1})\cos \theta_{1} \frac{\omega_{\rm p}}{n_{1}}-f_{3}(e_{1})\right),\\
\label{eq:edot}
\dot{e}_1&=&9\frac{\tilde{K_1}}{a_{1}^{8}}\left(\frac{11}{18}f_{4}(e_{1})\cos \theta_{1} \frac{\omega_{\rm p}}{n_{1}}-f_{5}(e_{1})\right)e_{1}, \end{eqnarray}
where $\theta_1$ is the angle between the spin of the inner planet and the inner orbit, $n_1$ is mean motion of the inner planet, and 
\begin{eqnarray}
\tilde{K_1}&=&\Delta{t_{\rm p}}\frac{3k_{2,1}Gm_{0}(m_{0}+m_{1})R_{1}^{5}}{m_{1}},\\
 f_{2}(e) &=& \frac{1+15e^{2}/2+45e^{4}/8+5e^{6}/16}{(1-e^{2})^{6}},\\
 f_{3}(e) &=& \frac{1+31e^{2}/2+255e^{4}/8+185e^{6}/16+25e^{8}/64}{(1-e^{2})^{15/2}},\\
 f_{4}(e) &=& \frac{1+3e^{2}/2+e^{4}/8}{(1-e^{2})^{5}},\\
 f_{5}(e) &=& \frac{1+15e^{2}/4+15e^{4}/8+5e^{6}/64}{(1-e^{2})^{13/2}}, 
\end{eqnarray}
where $\Delta {t_{\rm p}}$ is the tidal delay time of the inner planet which is related to the viscous time scale by $t_{\rm v,p}=3(1+k_{2,1})R_{1}^3/(Gm_1\Delta t_{\rm p})$. 

Because of the quick tidal realignment due to efficient planetary tides, the spin-orbit angle of the inner planet, $\theta_1$, effectively vanishes, and the spin rate of the inner planet, $\omega_{\rm p}$, reaches the following equilibrium state in a very short time scale \citep{Correia2011}:
\begin{eqnarray}  
\frac{\omega_{\rm p}}{n_{1}} = \frac{f_{2}(e_{1})}{f_{1}(e_{1})}, 
\label{eq:f1} \end{eqnarray}
where
\begin{eqnarray}  
f_{1}(e) &=& \frac{1+3e^{2}+3e^{4}/8}{(1-e^{2})^{9/2}}.
\label{eq:f1} \end{eqnarray}
Therefore, equations (\ref{eq:adot}) and (\ref{eq:edot}) reduce to the following simpler form
\begin{eqnarray}  
\dot{a}_1&=&2\frac{\tilde{K_1}}{a_{1}^{7}}\left(\frac{f_{2}^2(e_{1})}{f_{1}(e_{1})}-f_{3}(e_{1})\right), \label{eq:adot-r}\\
\dot{e}_1&=&9\frac{\tilde{K_1}}{a_{1}^{8}}\left(\frac{11}{18}\frac{f_{4}(e_{1})f_{2}(e_{1})}{f_{1}(e_{1})}-f_{5}(e_{1})\right)e_{1}. \label{eq:edot-r} \end{eqnarray}
Nevertheless, equations (\ref{eq:adot-r}) and (\ref{eq:edot-r}) cannot be yet solved analytically. In order to analytically estimate the migration time scale, we further assume that the migration process may be divided into two stages, eccentricity oscillation and circularization, based on the orbital evolution of the inner planet. 

Figure \ref{fig:ple94prosample1}a plots a typical example of the dynamical evolution near the lower boundary between NM and PHJ, corresponding to point A in Figure \ref{fig:plfid}. During the eccentricity oscillation ($t \leq 3\times10^8$yrs), the maximum eccentricity of the inner planet in each cycle, $e_{1,\rm max}$, is approximately constant against time, while the amplitude of the eccentricity variation damps. Eventually, the eccentricity oscillation stops ($t\geq 3\times10^8$ yrs) and the inner planet starts its circularization stage. 

To distinguish the behavior of the two different stages, we introduce the dimensionless orbital angular momentum of the inner planet: 
\begin{eqnarray}
\frac{p}{p_i} = \sqrt{\frac{a_{1}(1-e_{1}^{2})}{a_{1,i}(1-e_{1,i}^{2})}}, \label{eq:t-o1} 
\end{eqnarray}
which is plotted in the third panel of Figure \ref{fig:ple94prosample1}a. It indicates that the eccentricity oscillation stage also corresponds to the oscillation of $p/p_{i}$, where $p/p_{i}$ is constant during the circularization stage. 

Now, the migration time scale is simply approximated by the sum of the time scales of the above two stages:
\begin{eqnarray}
t_{\rm mig,1}=t_{\rm eo}+t_{\rm cir}.\label{eq:t-all} 
\end{eqnarray}
We present how to compute $t_{\rm eo}$ and $t_{\rm cir}$ separately below.

We first consider $t_{\rm eo}$. In the eccentricity oscillation stage, $e_{1,\rm max}$ is constant and can be computed from equation (\ref{eq:petrovich19}). The minimum eccentricity of the inner planet in each cycle, $e_{1,\rm min}$, steadily increases. The end of the eccentricity oscillation stage is the epoch that the amplitude of the eccentricity variation approximately vanishes. In order to separate the two stages, we denote $e_{1, \rm min, crit}$ as the value of $e_{1,\rm min}$ at the end of the eccentricity oscillation stage. According to \citet{Anderson2016}, we rewrite $e_{1,\rm min}$ in terms of $\Delta{j}$, where
\begin{eqnarray} 
\label{eq:delta-j} 
\Delta{j}=j_{1,\rm min}-j_{1,\rm max}=\sqrt{1-e_{1,\rm min}^{2}}-\sqrt{1-e_{1,\rm max}^{2}}.
\end{eqnarray}
Then we attempt to parameterize $e_{1,\rm min, crit}$ by assuming a constant critical value of $\Delta{j}_{\rm crit}$ as follows:
\begin{eqnarray}  
e_{1,\rm min,crit} = \sqrt{1-\left(\Delta{j}_{\rm crit} + \sqrt{(1-e_{1,\rm max}^{2})}\right)^2}.\label{eq:edot-e1min} \end{eqnarray}

Once $e_{1,\rm min, crit}$ and $e_{1,\rm max}$ are specified, the semi-major axis of the inner planet at the end of its eccentricity oscillation stage, $a_{1,\rm crit}$, is obtained from equation (\ref{eq:petrovich19}), combined with the conservation of the potential energy during each eccentricity oscillation cycle.

Using this $a_{1,\rm crit}$, $t_{\rm eo}$ is written  as
\begin{eqnarray}  
\label{eq:t-o1}
t_{\rm eo}= 
\int^{t(a_{1, \rm crit})}_{t(a_{1,i})} \, dt
\end{eqnarray}
The integral in the right-hand-side of the above equation is estimated as follows. 

During the eccentricity oscillation cycle, the damping of
$a_1$ due to the tidal interaction happens only for a short period when
$e_1$ is close to $e_{1,\rm max}$. Except for that period, $a_1$ is
approximately constant.  \citet{Anderson2016} found that the fraction of
the time that the inner planet is in the high-$e$ phase ($e_1 \approx
e_{1,\rm max}$) in each eccentricity oscillation cycle is given by
$\sqrt{1-e_{1,\rm max}^2}$. In reality they derived the above result in the Lidov-Kozai regime, but it is also applicable for the near-coplanar configuration that we consider here. Therefore $t_{\rm eo}$ should be givenas
\begin{eqnarray}
t_{\rm eo} = \frac{\Delta t(e_{1, \rm max})}{\sqrt{1-e_{1,\rm max}^2}}.
\end{eqnarray}
where $\Delta t(e_{1, \rm max})$ is the total duration when the inner planet has $e_1 \approx e_{1,\rm max}$ during the entire eccentricity oscillation stage:
\begin{eqnarray}  
\label{eq:Deltat-e}
\Delta t(e_{1, \rm max})
\approx
\int^{a_{1,i}}_{a_{1, \rm crit}} 
\left|\frac{dt}{da_1}\right|_{e_1=e_{1,\rm max}} \, da_1 .
\end{eqnarray}
Equation (\ref{eq:Deltat-e}) is further approximated and evaluated using equations (\ref{eq:adot-r}):
\begin{eqnarray}  
\Delta t(e_{1, \rm max})
&\approx&
\int^{a_{1,i}}_{a_{1, \rm crit}} 
\left|\frac{dt}{da_1}\right|_{e_1=e_{1,\rm max}}da_1 \cr
&=&
\int^{a_{1,i}}_{a_{1, \rm crit}}
\frac{a_{1}^{7}}{2\tilde{K_1}}\left(\frac{f_{2}^2(e_{1,\rm max})}{f_{1}(e_{1,\rm max})}-f_{3}(e_{1,\rm max})\right)^{-1} \cr
 &=&\frac{a_{1,i}^8-a_{1,\rm crit}^8}{16\tilde{K_1}}\left(\frac{f_{2}^2(e_{1,\rm max})}{f_{1}(e_{1,\rm max})}-f_{3}(e_{1,\rm max})\right)^{-1}. 
\label{eq:t-eo}
\end{eqnarray}

Next, we move to the calculation of $t_{\rm cir}$. The equilibrium value of the dimensionless orbital angular momentum of the inner planet in the circularization stage, $p_{\rm eq}/p_i$, can be obtained at the end of the eccentricity oscillation stage as 
\begin{eqnarray}
\frac{p_{\rm eq}}{p_i}=\sqrt{\frac{a_{1,\rm crit}(1-e_{1,\rm max}^{2})}{a_{1,i}(1-e_{1,i}^{2})}}. \label{eq:a_feq} 
\end{eqnarray}
Then, we define the end of the circularization stage when the inner planet becomes a HJ with $e_{1}<0.01$. Thus, from equation (\ref{eq:edot-r}), $t_{\rm cir}$ can be written as 
\begin{eqnarray}
t_{\rm cir}&=& \int^{e_{1,\rm max}}_{0.01} \left|\frac{dt}{de_1}\right|_{a_1=p_{\rm eq}^2/(1-e_{1}^{2})}de_1\nonumber\\
 &=&\int^{e_{1,\rm max}}_{0.01} \frac{p_{\rm eq}^{16}}{9\tilde{K_1}}\frac{1}{e_{1}(1-e_{1}^{2})^8}\left(\frac{11}{18}\frac{f_{4}(e_{1})f_{2}(e_{1})}{f_{1}(e_{1})}-f_{5}(e_{1})\right)^{-1}de_1.\label{eq:t-cir} 
\end{eqnarray}

The migration boundary is determined by the epoch when $t_{\rm mig,1}$ is equal to $T_{\rm max}$ ($10^{10}$ yrs in our simulations). The resulting lower and upper migration boundaries of $\epsilon$ in the fiducial case from equations (\ref{eq:t-all}), (\ref{eq:t-eo}), and (\ref{eq:t-cir}) are shown in Figure \ref{fig:plfid} by magenta lines. We observe that they are in good agreement with the numerical simulations. This analytical estimate is useful in interpreting the present simulation results. In addition, it provides a useful guidance for future numerical simulations in near-coplanar hierarchical triple systems and the same result is also applicable to the Lidov-Kozai regime. 

In Figure \ref{fig:plfid} and also Figure \ref{fig:5mjmdep2} below, we plot the boundary corresponding to $\Delta{j}_{\rm crit}=0.04$. Indeed, the results turn out to be fairly insensitive to the value of $\Delta{j}_{\rm crit}$ in a certain range. In the upper panel of Figure \ref{fig:ple94prosample1}a, the black, magenta, and blue vertical lines show the analytical estimates for $t_{\rm eo}$ and $t_{\rm mig,1}$ with $\Delta{j}_{\rm crit}=0.03,0.04$, and $0.05$, respectively. We observe that the smaller $\Delta{j}_{\rm crit}$ implies a longer time scale of the eccentricity oscillation stage and a shorter time scale of the circularization stage. The dependence of $\Delta{j}_{\rm crit}$ on the migration time scale is determined by the above two competitive effects. Nevertheless, we confirm that adopting different $\Delta{j}_{\rm crit}$ does not qualitatively change the estimated time scale.

\subsubsection{Stopping Time of NM and PHJ\label{sec:ts-nmphj}}

\begin{figure}[t]
\begin{center}
\includegraphics[width=16cm]{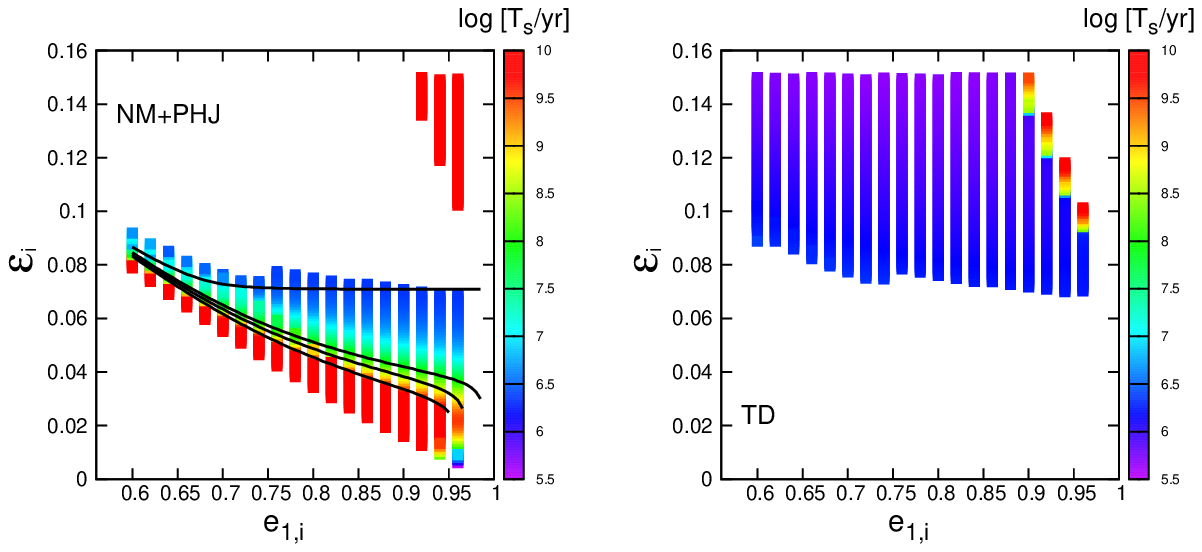} 
\caption{Stopping time, $T_{\rm s}$, time at which each simulation is stopped is plotted in log scale in the fiducial case. $T_{\rm s}$ is defined as $T_{\rm max}$ for NMs, the migration time scale for PHJs, and the time scale to reach less than the Roche limit for TDs. The left panel refers to $T_{\rm s}$ of NMs and PHJs. The right panel presents $T_{\rm s}$ of TDs. The black solid lines in the left panel correspond to our analytical estimate of the migration time scale, $10^{10}$ yr, $10^{9}$ yr, $10^{8}$ yr, and $10^{7}$ yr, from bottom to top.} \label{fig:time-plot}
\end{center}
\end{figure}

The stopping time, $T_{\rm s}$, the time at which each simulation is stopped, provides an important hint on the stability of the system. Figure \ref{fig:time-plot} presents the stopping time of NM and PHJ (left), and TD (right) corresponding to our fiducial runs in Figure \ref{fig:plfid}. In this subsection, we discuss the stopping time of NM and PHJ.

Since NMs stay almost at the initial position until the maximum simulation time, $T_{\rm max}$, the stopping time of NM is simply $T_{\rm max}$ that we adopt. The stopping time of PHJ is determined by the epoch when the inner planet satisfies $a_{1,f}<0.1$ AU and $e_{1,f}<0.01$ simultaneously, which is equal to the migration time scale. As shown in the left panel of Figure \ref{fig:time-plot}, the stopping time of PHJ almost monotonically decreases from $T_{\rm max}=10^{10}$ yr to $\sim10^6$ yr as $\epsilon_i$ increases. 

Depending on the migration time scale, we adopt two different analytical approaches to understand the above behavior. For $T_{\rm s}=10^{10}$ yr and $10^{9}$ yr, we use equation (\ref{eq:t-all}), but for $T_{\rm s}=10^{8}$ yr and $10^{7}$ yr, we have to use a different approach because the orbit evolves in a different manner than in the previous case. In PHJ region, as $\epsilon_i$ increases, the path to PHJ happens over a much smaller time scale. The time evolution for an example of PHJ systems in that region, corresponding to point B in Figure \ref{fig:plfid}, is plotted in Figure \ref{fig:ple94prosample1}b with $T_{\rm s}\sim 10^7$ yr. As shown in its second panel, $1-e_1$ monotonically decreases and reaches less than $10^{-2}$, when tides dominate the orbital evolution of the inner planet and circularize its orbit within several million years. During the circularization stage, the inner orbital angular momentum is conserved. Unlike the systems near the lower migration boundary (Figure \ref{fig:ple94prosample1}a), the eccentricity oscillation does not happen in this region. Thus, the migration time scale becomes the sum of the time scales of the first eccentricity growth, $t_{\rm eg}$, and subsequent tidal circularization with the constant inner orbital angular momentum. Following \citet{Petrovich2015b}, we compute $t_{\rm eg}$ from his $\tau_{\rm in}/\alpha$:
\begin{eqnarray}
t_{\rm eg}=\frac{4}{3n_1}\left(\frac{m_0}{m_2}\right)\left(\frac{a_2}{a_1}\right)^4.\label{eq:teg} 
\end{eqnarray}
The circularization time scale can be computed from equation (\ref{eq:t-cir}). Then, the total migration time scale in this region is given by
\begin{eqnarray}
\label{eq:decay_short} 
t_{\rm mig,2}=t_{\rm eg}+t_{\rm cir}.\label{eq:t-all-2} 
\end{eqnarray}
The left panel of Figure \ref{fig:time-plot} plots the analytical estimate of the four migration time scales: for $10^{10}$ yr and $10^{9}$ yr, we use equation (\ref{eq:t-all}), and for $10^{8}$ yr and $10^{7}$ yr, we adopt equation (\ref{eq:t-all-2}). Combining above two methods, we find that the results are roughly consistent with our numerical simulation.

\subsubsection{Stopping Time of TD \label{sec:ts-td}}

The stopping time of TD is determined by the epoch when $q_1$ reaches less than the Roche limit. The results corresponding to the fiducial runs in Figure \ref{fig:plfid} are shown in the right panel of Figure \ref{fig:time-plot}. It implies the stopping time of the majority of TDs is $\sim10^6$ yr, indicating that these planets fall into the Roche limit at the first few extreme eccentricity approaches \citep{XueSuto2016}. In the region close to the upper migration boundary, $T_{\rm s}$ becomes longer as $\epsilon_i$ increases. The time evolution for an example of TD systems in that region, corresponding to point C in Figure \ref{fig:plfid}, is shown in Figure \ref{fig:ple94prosample1}c. This system has $T_{\rm s}\sim10^9$ yr. As shown in its bottom panel, $|\omega_1-\omega_2|$ librates with decreasing amplitude, and then starts circulating at $\sim7\times10^8$ yr. Exactly at the transition between libration and circulation, the inner planet acquires extreme eccentricity and therefore becomes TD. 

\subsection{Fate of the Inner Planet in Non-fiducial Models\label{sec:sim-dep}}

\begin{figure}[t]
\begin{center}
\includegraphics[width=18cm]{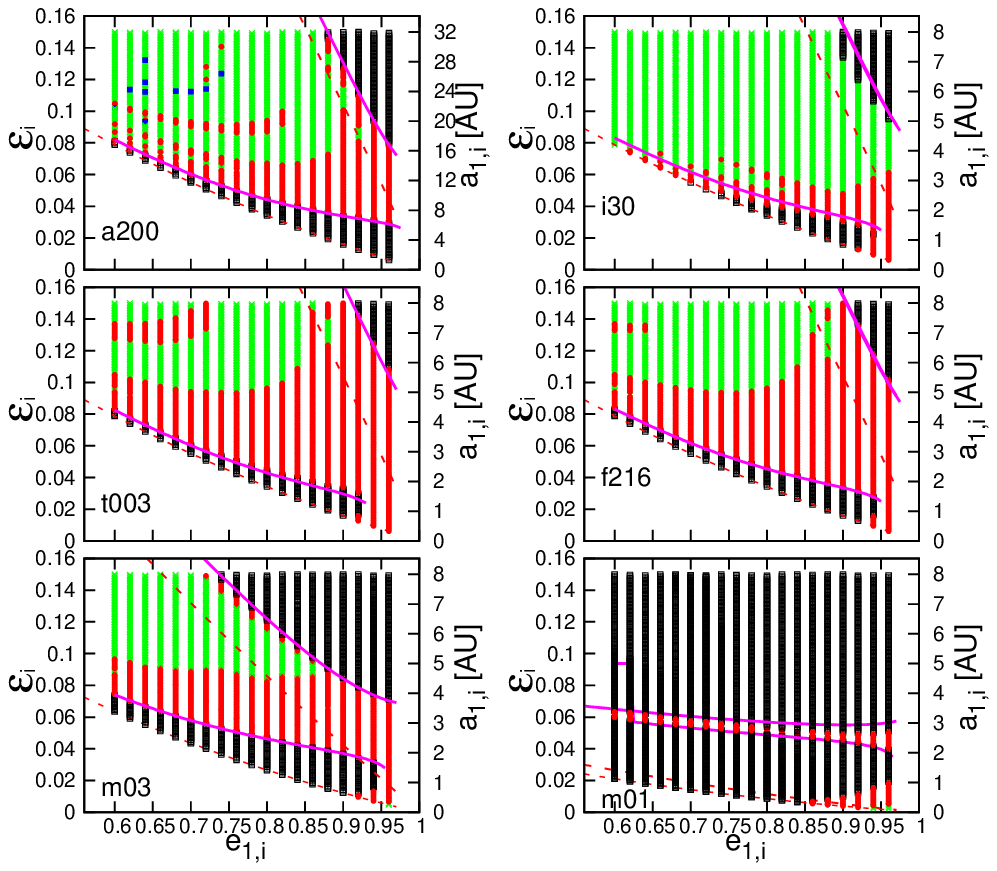} 
\caption{Fate of the inner planet on ($e_{1,i},\epsilon_{i}$) plane for six selected sets, left upper: a200 $(a_{2,i}=200$ AU); right upper: i30 ($i_{12,i}=30^{\circ}$); left middle: t003 ($t_{\rm v,p}=0.03$ yr); right middle: f216 ($f=2.16$); left bottom: m03 ($m_2=3M_{\rm J}$); and right bottom: m01 ($m_2=1M_{\rm J}$), respectively. The red dashed lines show the extreme eccentricity condition derived from equation (\ref{eq:epsilon-oct}). The magenta solid lines correspond to the analytical estimate of the migration boundary obtained from equation (\ref{eq:t-all}) with $\Delta j_{\rm crit} = 0.04$.} \label{fig:5mjmdep2}
\end{center}
\end{figure}

The previous subsections presented the simulation result and its implications in the fiducial case. Next we consider the parameter dependence of the final outcomes. The results on $(e_{1,i},\epsilon_i)$ plane for selected six different cases are plotted in Figure \ref{fig:5mjmdep2} (see also Table \ref{tab:pl}). 

The parameter dependence can be understood by comparing each model with the fiducial case. In the case a200 ($a_{2}=200$ AU), shown in the left upper panel of Figure \ref{fig:5mjmdep2}, the fraction of NMs increases since for the same $\epsilon_i$, a larger $a_{1,i}$ corresponds to a larger $q_{1,\rm min}$. The right upper panel of Figure \ref{fig:5mjmdep2} plots the case i30 ($i_{12,i}=30^{\circ}$). The fraction of TDs increases since the larger amplitude of $i_{12,i}$ oscillation induces a higher maximum achievable eccentricity of the inner planet. In the case t00030 ($t_{\rm v,p}=0.003$ yr), shown in the left middle panel of Figure \ref{fig:5mjmdep2}, the stronger tides on the inner planet cause it to suffer from the very efficient tidal dissipation even at a relatively larger pericenter distance, resulting in a higher fraction of PHJs. The middle right panel of Figure \ref{fig:5mjmdep2} corresponds to the case f216 ($f=2.16$), in which the smaller disruption radius results in less TDs and more PHJs.

Such dependence on $m_2$, $i_{12,i}$, $t_{\rm v,p}$, and $f$ is similar to the sub-stellar perturber case considered in \citet{XueSuto2016}, but the dependence on $m_2$ generates a different feature. Thus, in this subsection, we mainly focus on the dependence on $m_2$. Here we restrict our attention to the planetary perturber. We perform eight different sets of runs with $m_{2}=10M_{\rm J}$, $7M_{\rm J}$, $6M_{\rm J}$, $5M_{\rm J}$ (fiducial), $4M_{\rm J}$, $3M_{\rm J}$, $2M_{\rm J}$, and $1M_{\rm J}$. The final states of the inner planet for those models are summarized in Table \ref{tab:pl}. Also, two examples with $m_2=3M_{\rm J}$ and $1M_{\rm J}$ on $(e_{1,i},\epsilon_i)$ plane are plotted in the left and right bottom panels of Figure \ref{fig:5mjmdep2}. 

Table \ref{tab:pl} shows the fraction of NMs increases as $m_2$ decreases. As shown in the right panel of Figure \ref{fig:flip1}, the extreme eccentricity region becomes narrower as $m_2$ decreases. In addition, the relative importance of GR increases as $m_2$ decreases according to equation (\ref{eq:GR-pre}). Therefore the short-range forces more efficiently limit the extreme eccentricity growth for smaller $m_2$. The above two facts account for the anti-correlation between the fraction of NMs and the value of $m_2$.

The fraction of TDs decreases as $m_2$ decreases, which can be understood as follows. The orbital interaction is the major driving source for the inner planet acquiring the extreme eccentricity. The weaker interaction (smaller $m_2$) leads to the smaller maximum achievable eccentricity of the inner planet, and therefore to the larger $q_{1,\rm min}$. As a result, fewer systems suffer from the tidal disruption. 

As $m_2$ decreases, the fraction of NMs increases but that of TDs decreases. Thus the change of the fraction of PHJs depends on these competitive effects. In our simulations, the fraction of PHJs increases as $m_2$ decreases from $10M_{\rm J}$ to $2M_{\rm J}$, and even more significantly decreases as $m_2$ decreases from $2M_{\rm J}$ to $1M_{\rm J}$. In the latter case, the migration region becomes significantly narrower as shown in the right bottom panel of Figure \ref{fig:5mjmdep2}. Most of simulation runs end up with NMs. The above mentioned trends suggest that in order to form HJ in CHEM, the outer perturber should be neither too small to over-limit the migration region, nor too large to be dominated by disruption. An intermediate massive perturber is preferred.

\begin{figure}[t]
\begin{center}
\includegraphics[width=18cm]{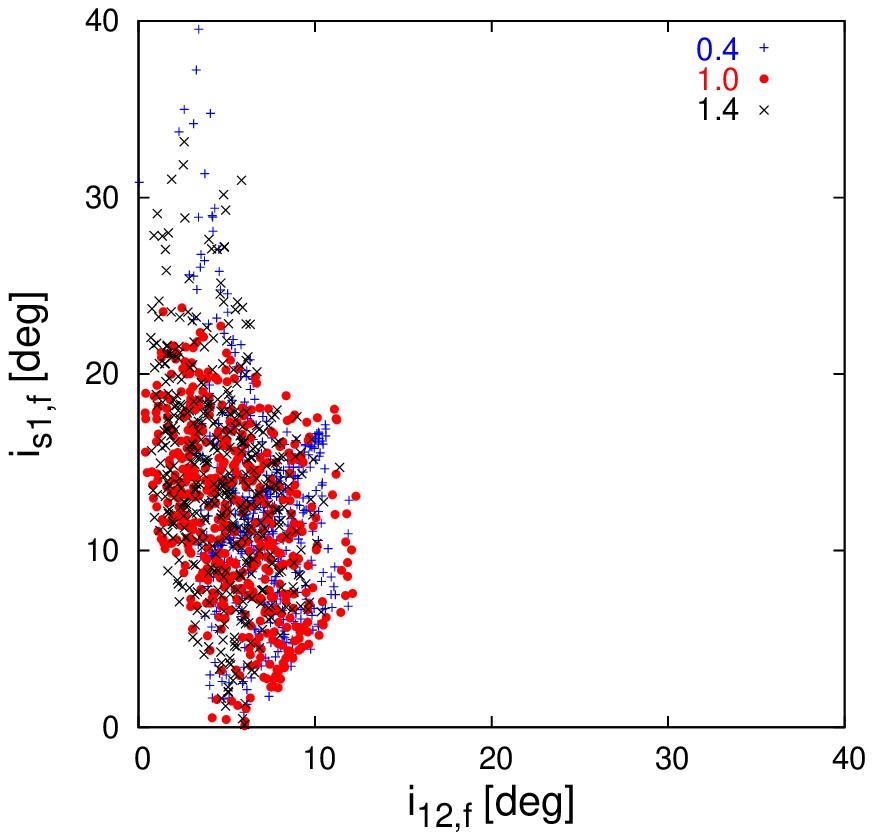} 
\caption{Orbital mutual orbital inclination, $i_{12,f}$, against the spin-orbit angle between the central star and the inner planet, $i_{s1,f}$, for resulting PHJs among $m_0=0.4M_{\odot}$ (blue plus), the fiducial case ($m_0=1M_{\odot}$, red dot), and $m_0=1.4M_{\odot}$ (black cross).} \label{fig:m0depis1i12}
\end{center}
\end{figure}

Recently, \citet{Anderson2016} examined the dependence of the final spin-orbit angle on the mass of the central star in the Lidov-Kozai migration with a circular outer perturber. They found that the systems with more massive central stars have broader distribution of the spin-orbit angle. Thus, we attempt to see the possible dependence of the spin-orbit angle on the mass of the central stars in CHEM. The comparison of cases $m_0=0.4M_{\odot}$ and $1.4M_{\odot}$ cases against our fiducial case ($m_0=1M_{\odot}$) indicates very similar distribution for the spin-orbit angle as shown in Figure \ref{fig:m0depis1i12}. This is indeed consistent with \citet{XueSuto2016}; see their Figure 12. Therefore, the spin-orbit angle in CHEM does not seem to be sensitive to the mass of the central star.

\subsection{Distribution of Final Semi-major Axis $a_{1,f}$\label{sec:af-dis}}

\begin{figure}[t]
\begin{center}
\includegraphics[width=18cm]{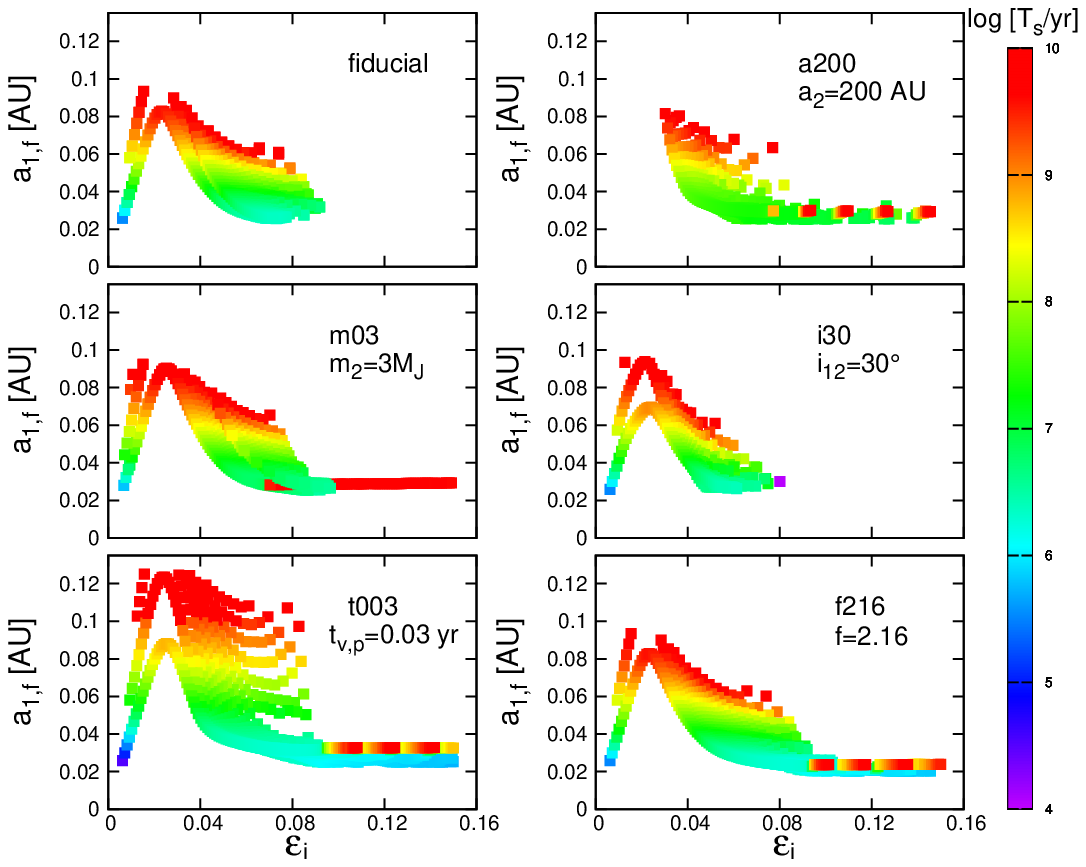} 
\caption{Final semi-major axis of resulting HJs, $a_{1,f}$, of six selected cases against $\epsilon_i$, left upper: fiducial; right upper: a200 $(a_{2,i}=200$ AU); left middle: m03 ($m_2=3M_{\rm J}$); right middle: i30 ($i_{12,i}=30^{\circ}$); left bottom: t003 ($t_{\rm v,p}=0.03$ yr); and right bottom: f216 ($f=2.16$), respectively. The different colors indicate the different stopping time $T_{\rm s}$ in log scale.} \label{fig:af-dep}
\end{center}
\end{figure}

The final distribution of the orbital elements in our simulations provide possible hints in distinguishing CHEM from the other HJ formation mechanisms. In this subsection, we discuss the distribution and its parameter dependence of the final semi-major axis of the resulting HJs, $a_{1,f}$, in our simulations. Figure \ref{fig:af-dep} presents $a_{1,f}$ against $\epsilon_i$ in six selected simulation sets. The colors represent the stopping time $T_{\rm s}$ in log scale. Note, in the case of PHJs, $T_{\rm s}$ is equivalent to their migration time scale. 

We plot $a_{1,f}$ in the fiducial case in the left upper panel of Figure \ref{fig:af-dep}. There are mainly two important features. One is that $a_{1,f}$ is distributed with $\sim0.025-0.096$ AU, which is qualitatively consistent with the observation. The other is that $T_{\rm s}$ basically increases as $a_{1,f}$. The detailed explanation of the above two features is presented below.

In the fiducial case, the lower boundary of $a_{1,f}\sim0.025$ AU is roughly consistent with twice the Roche limit $\sim0.0126$ AU. It comes from the fact that $a_{1,f}\approx2q_{1,\rm min}$ during tidal circularization due to the constant inner orbital angular momentum as described in subsection \ref{sec:mig-time}. Here $q_{1,\rm min}$ of the resulting HJs is larger than the Roche limit in order to survive the disruption. In the case of f216 ($f=2.16$) with less restrictive disruption radius shown in the bottom right panel of Figure \ref{fig:af-dep}, the lower boundary of $a_{1,f}$ reduces to $\sim0.02$ AU. The upper boundary of $a_{1,f}$ increases as the efficiency of tides, since the planetary orbit is circularized within $T_{\rm max}$ even at a larger distance if tides are stronger. In the case of t00030 ($t_{\rm v,p}=0.003$ yr) with stronger tides shown in the left bottom panel of Figure \ref{fig:af-dep}, the upper boundary is extended to $\sim0.13$ AU. Compared with the fiducial case, the range of $a_{1,f}$ is fairly insensitive to the change of the parameters of the outer perturber, $a_2$, $m_2$, and $i_{12}$, as shown in the right upper, the left middle, and the right middle panels of Figure \ref{fig:af-dep}, respectively.

The trend that $T_{\rm s}$ (also the migration time scale) basically increases as $a_{1,f}$ holds for all the cases. This trend can be explained by the correlation between $q_{1,\rm min}$ and the migration time scale. During the tidal circularization, the correlation $a_{1,f}\approx2q_{1,\rm min}$ holds, where $q_{1,\rm min}$ is very sensitive to the strength of tides, and therefore to the migration time scale. This trend is broken for the systems near the upper migration boundary, which are located in $\epsilon_i>0.1$ as shown in Figure \ref{fig:af-dep}. In that region, the systems take a longer time to reach $q_{1,\rm min}$. This is the case for point C in Figure \ref{fig:plfid}, and its dynamical behavior is plotted in Figure \ref{fig:ple94prosample1}c. The system reaches $q_{1,\rm min}$ at the transition epoch of $|\omega_1-\omega_2|$ from libration to circulation as discussed in subsection \ref{sec:ts-td}. Since a smaller amplitude of libration for $|\omega_1-\omega_2|$ corresponds to a stronger modulated envelope of $e_1$, the time spent near $q_{1,\rm min}$ becomes smaller to avoid fast circularization or disruption \citep{Dawson2014}.

\section{Super-Earth with a Planetary Outer Perturber\label{sec:superearth}}

About forty close-in super-Earths ($a_1<0.1$ AU) have been observed by Kepler so far. The origin of those planets is also an open question similarly as HJs. It was proposed that close-in super-Earths may form in situ or by disk migration, but neither of them can fully explain the current observation. The former scenario requires that the proto-planetary disk should be at least 20 times more massive than that in the minimum-mass solar nebula \citep{Raymond2014}. The latter predicts that the multi-planets should be in low-order mean motion resonance, but it is not supported by the observation \citep{Chatterjee2015}. Recently, \citet{Rice2015} suggested that the ultrashort-period super-Earth Kepler-78b with $a=0.009$ AU may be explained by the Lidov-Kozai migration, which implies that the dynamical processes may also contribute to the formation of close-in super-Earths at least partially. In this section, therefore, we examine to see if CHEM can produce close-in super-Earths. To be more specific, we consider an inner super-Earth at $\sim1$ AU and an outer eccentric planetary perturber at $\sim10$ AU initially, because super-Earths are almost preferentially found within the snow line from the observation. Our initial condition differs from \citet{Rice2015} in considering the planetary perturber instead of the stellar perturber, and such a configuration may result from planet-planet scattering. 

\begin{figure}[t]
\begin{center}
\includegraphics[width=18cm]{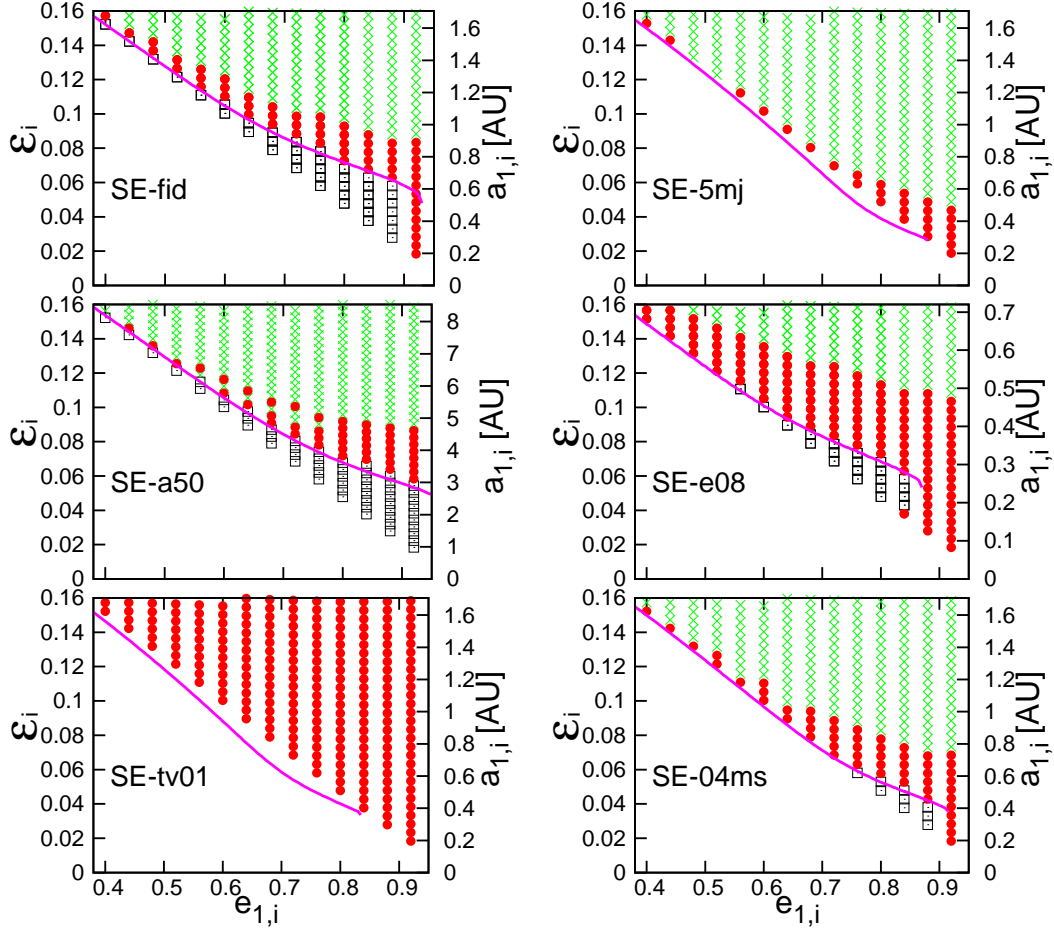} 
\caption{Fate of the inner planets on ($e_{1,i},\epsilon_{i}$) plane for six different cases in super-Earth system simulations, SE-fid, SE-5mj, SE-a50, SE-e08, SE-tv01, and SE-04ms. The final states are indicated by black open squares for non-migrating planets (NM), red filled circles for prograde super-Earth (PSE), green crosses for tidally disrupted planets (TD), respectively. The magenta solid lines correspond to the migration boundary $t_{\rm mig,1} = 10^{10}$ yr based on the analytical estimate derived from equation (\ref{eq:t-all}).} \label{fig:superearth}
\end{center}
\end{figure}

We perform the numerical simulations following the procedure described in section \ref{sec:pl}, but we change the inner planetary mass and radius, tidal strength, and disruption radius in order to adjust to the current situation. We adopt initially $m_{1} = 5M_{\oplus}$, $i_{12}=6^{\circ}$, $i_{s1} = 0^{\circ}$, $\omega_{1} = 0^{\circ}$, $\omega_{2} = 0^{\circ}$, $\Omega_{1} = 180^{\circ}$, $\Omega_{2} = 0^{\circ}$, and $f=2.44$ for all the models. The radius of the inner planet is determined by the planet mass-radius relationship: $M/M_{\oplus} = 2.69(R/R_{\oplus})^{0.93}$ \citep{Weiss2014}. We perform $\sim$ 300 different runs by systematically varying $(e_{1,i},\epsilon_i)$. $e_{1,i}$ is varied between $0.6$ and $0.96$ with a constant interval of $0.04$, and  $\epsilon_i$ is varied between $\epsilon_{\rm L}$ derived from equation (\ref{eq:epsilon-oct}) and $0.16$ with a constant interval of $0.004$. The other parameters for six different models are summarized in Table \ref{tab:superearth} together with the final fraction of the different outcomes. The resulting $(e_{1,i},\epsilon_i)$ maps are presented in Figure \ref{fig:superearth}. We first discuss the result in the fiducial case, and then consider the dependence on $m_2$, $a_2$, $e_2$, $t_{\rm v,p}$, and $m_0$. 

In the fiducial case, SE-fid, we adopt $m_0=1.0M_{\odot}$, $a_2=10$ AU, $e_2=0.6$, $m_2=1M_{\rm J}$, and $t_{\rm v,p}=0.001$ yr. The viscous time scale of the inner planet, $t_{\rm v,p}=0.001$ yr, is taken from the value of Earth \citep{Murray1999}, which corresponds to the quality factor $Q\sim100$ for a 1 yr orbital period. The resulting $(e_{1,i},\epsilon_i)$ map of the fiducial case is shown in the left upper panel of Figure \ref{fig:superearth}. Clearly, the overall distribution is very similar to the case of a giant gas inner planet with a planetary outer perturber as shown in Figure \ref{fig:plfid}. The final outcomes are NM (19.9\%), PSE (22.2\%), and TD (57.9\%), where PSE refers to the prograde close-in super-Earth ($a_{1,f}<0.1$ AU, $i_{12,f}<90^{\circ}$). We note that we do not find any retrograde close-in super-Earth (RSE, $a_{1,f}<0.1$ AU, $i_{12,f}>90^{\circ}$). The absence of RSE is supposed to be generic because the initial location of super-Earth is likely within the snow line, and therefore tides circularize the orbit before super-Earth acquires the extreme eccentricity necessary for the orbital flip. The migration boundary between NM and PHJ for $t_{\rm mig,1}=10^{10}$ yr based on equation (\ref{eq:t-all}) is plotted in magenta line, which is in good agreement with our numerical simulation.

Next, we consider the parameter dependence by comparing with the fiducial case. We find that the dependence on $m_2$, $a_{2}$, $e_{2,i}$, and $t_{\rm v,p}$ are similar to the cases for systems consisting of a giant gas inner planet with a sub-stellar outer perturber \citep{XueSuto2016} and with a planetary outer perturber in section \ref{sec:pl}. In the case SE-5mj ($m_2 = 5M_{\rm J}$) shown in the right upper panel of Figure \ref{fig:superearth}, decrease of NM and increase of TD are due to the stronger mutual orbital interaction that leads to a more extreme eccentricity. In the case SE-a50 ($a_2 = 50$ AU) plotted in the left middle panel of Figure \ref{fig:superearth}, NM increases because a larger $a_{1,i}$ corresponds to a larger $q_{1,\rm min}$. The right middle panel of Figure \ref{fig:superearth} refers to the case SE-e08 ($e_{2,i} = 0.8$). PSE increases due to the smaller $a_{1,i}$ by scaling law according to equation (\ref{eq:eoct}). In the case SE-tv01 ($t_{\rm v,p} = 0.00001$ yr) shown in the left bottom panel of Figure \ref{fig:superearth}, more systems survive as PSEs due to very strong tides.

In the case of simulations for super-Earth systems, we additionally consider the case with smaller $m_0$, because super-Earths are often found around M-dwarfs. The related case SE-04ms is shown in the right bottom panel of Figure \ref{fig:superearth}, where we decrease the mass of the central star to $m_0=0.4M_{\odot}$. In this case, PSEs have a similar fraction compared with the fiducial case, but form in the lower $\epsilon_i$ region. The fraction of NM decreases from 19.9\% in the fiducial case to 4.4\%. This trend is due to decrease of the relative importance of the short-range force effect. The potential energy of GR ($\tilde{\phi}_{\rm GR}$) and tides ($\tilde{\phi}_{\rm tide}$) decrease with $m_0$, but that of orbital interaction ($\tilde{\phi}_{\rm}$) remains constant. Therefore the extreme eccentricity growth is less suppressed than in the fiducial case.

In summary, the overall distribution and parameter dependence in simulations are similar between super-Earth systems and giant gas planetary systems. No RSE is observed in our simulations, but CHEM can produce PSE to some degree. Thus, we conclude that CHEM is a possible channel to form close-in super-Earths.

\begin{table}[t]
\begin{center}
\caption{Summary of parameters and the facts of simulation runs described in section \ref{sec:superearth}.}
\begin{threeparttable}
\begin{tabular}{|c||c|c|c|c|c||c|c|c|c|}
\hline 
Model&$m_{0}$&$a_{2}$& $e_{2}$ & $m_{2}$& $t_{\rm v, \rm p}$ &PSE&RSE&NM&TD\\ 
&$M_{\odot}$&au&&$M_{\rm J}$&yr&&&& \\  \hline
SE-fid&1.0&$10$&0.6& 1 &0.001 &22.2\%&0.0\%&19.9\%&57.9\%\\\hline
SE-5mj&1.0&$10$&0.6& 5  & 0.001&12.5\%&0.0\%&0.0\%&87.5\%\\
SE-a50&1.0&$50$&0.6& 1  & 0.001&16.6\%&0.0\%&22.7\%&60.7\%\\ 
SE-e08&1.0&$10$&0.8& 1& 0.001&55.6\%&0.0\%&12.0\%&32.4\%\\
SE-tv01&1.0&$10$&0.6& 1 & 0.00001&100.0\%&0.0\%&0.0\%&0.0\%\\
SE-04ms&0.4&$10$&0.6& $1$ &0.001&23.9\%&0.0\%&4.1\%&72.0\%\\ \hline
\end{tabular}
\begin{tablenotes}[para,flushleft]
\item{\textbf{Note.} PSE and RSE refer to prograde close-in super-Earth ($a_{1,f}<0.1$ AU, $i_{12,f}<90^{\circ}$) and retrograde close-in super-Earth ($a_{1,f}<0.1$ AU, $i_{12,f}>90^{\circ}$), respectively. All the models adopt $m_{1} = 5M_{\oplus}$, $i_{12,i}=6^{\circ}$, $\omega_{1,i} = 0^{\circ}$, $\omega_{2,i} = 0^{\circ}$, $\Omega_{1,i} = 180^{\circ}$, $\Omega_{2,i} = 0^{\circ}$, $f=2.44$, and $i_{s1,i} = 0^{\circ}$. For each model, we perform $\sim300$ runs by varying $(e_{1,i}$, $\epsilon_{i})$ systematically.} 
\end{tablenotes}
\end{threeparttable}
\label{tab:superearth}
\end{center}
\end{table}

\clearpage

\section{Application to Observed Exoplanetary Systems \label{sec:real}}

So far, we have theoretically explored the fate of near-coplanar hierarchical triple systems and their parameter dependence in CHEM, but it has not been clear if this scenario can indeed explain the observed exoplanetary systems. In order to answer this question, we apply CHEM to two distinct groups of the observed exoplanetary systems: hierarchical triple and counter-orbiting HJ systems.

Table \ref{tab:data-obs} lists the observed parameters for the systems selected in our simulations. For hierarchical triple systems, we choose three close-in super-Earth systems, Kepler-93, Kepler-97, and Kepler-407 out of 22 Kepler systems in \citet{Marcy2014}. In addition, four HJ systems are selected from \citet{Knutson2014}. They reported possible outer companions for 14 HJ systems, and four of them, HAT-P-2, HAT-P-4, HAT-P-17, and WASP-22, have $\lambda<20^{\circ}$. Note that the true spin-orbit angle is larger than the projected one, $\lambda$, as illustrated in the footnote of section \ref{sec:intro}. For simplicity, however, we assume these four systems are well-aligned. Thus, we consider these four systems in our simulations as possible candidates for CHEM. Among the currently known HJ systems with measured $\lambda$, two of them are possibly counter-orbiting ($\lambda>160^{\circ}$), so we also consider these two systems; HAT-P-6 with $\lambda = 165^\circ \pm{6^\circ}$ \citep{Albrecht2012} and HAT-P-14 with $\lambda = 189^{\circ}.1\pm {5^\circ.1}$ \citep{Winn2011}. 

We fix the values of $m_0$, $m_1$, $m_2$, and $a_2$ as in Table \ref{tab:data-sim} for definiteness. We basically set the eccentricity of the outer perturber, $e_2$, as 0.6, and consider the dependence on $e_2$ in the case of Kepler-97 simulations. Following sections \ref{sec:pl} and \ref{sec:superearth}, we adopt initially $i_{12}=6^{\circ}$, $\omega_{1} = 0^{\circ}$, $\omega_{2} = 0^{\circ}$, $\Omega_{1} = 180^{\circ}$, $\Omega_{2} = 0^{\circ}$, and $i_{s1} = 0^{\circ}$ for all the systems. The viscous time scale of the inner planet $t_{\rm v,p}$ is set to 0.03 yr for HJ systems and 0.001 yr for super-Earth systems; the disruption control parameter $f$ is $2.7$ for HJ systems and $2.44$ for super-Earth systems, respectively. For all systems listed in Table \ref{tab:data-sim}, we survey ($e_{1,i},\epsilon_{i}$) plane similarly as in the previous sections; $e_{1,i}$ is varied between $0.4$ and $0.92$ (0.96) with a constant interval of $0.04$ (0.02), and  $\epsilon_i$ is varied between $\epsilon_{\rm L}$ derived from equation (\ref{eq:epsilon-oct}) and $0.15$ with a constant interval of $0.005$ (0.001) for hierarchical triple systems (counter-orbiting HJ systems).

We aim to examine if those observed systems can be reproduced by CHEM. For hierarchical triple systems, we require that the final semi-major axis of the inner planet, $a_{1,f}$, should be within the observed range. For counter-orbiting HJ systems, we further require that the final spin-orbit angle, $i_{s1,f}>160^{\circ}$, in addition to $a_{1,f}$ being within the observed range. Note, however, that the inner planets of HAT-P-2 and HAT-P-17 have finite eccentricities of $0.517$ and $0.346$, respectively. They are supposed to be still during the tidal circularization stage with a constant inner orbital angular momentum. Therefore, the final semi-major axis of the inner planet after the circularization stage should be $a_{\rm eq}=a_{1,\rm obs}(1-e_{1,\rm obs}^2)$, which is 0.0494 AU for HAT-P-2, and 0.0776 AU for HAT-P-17, respectively. We require that $a_{1,f}$ should be sufficiently close to $a_{\rm eq}$ for these two systems. 

We first focus on Kepler-97, a typical close-in super-Earth system, by running a variety of simulation models with different orbital parameters. This is also useful in understanding the parameter dependence in general.

\subsection{Kepler-97 \label{sec:real-k97}}

\begin{figure}[t]
\begin{center}
\includegraphics[width=18cm]{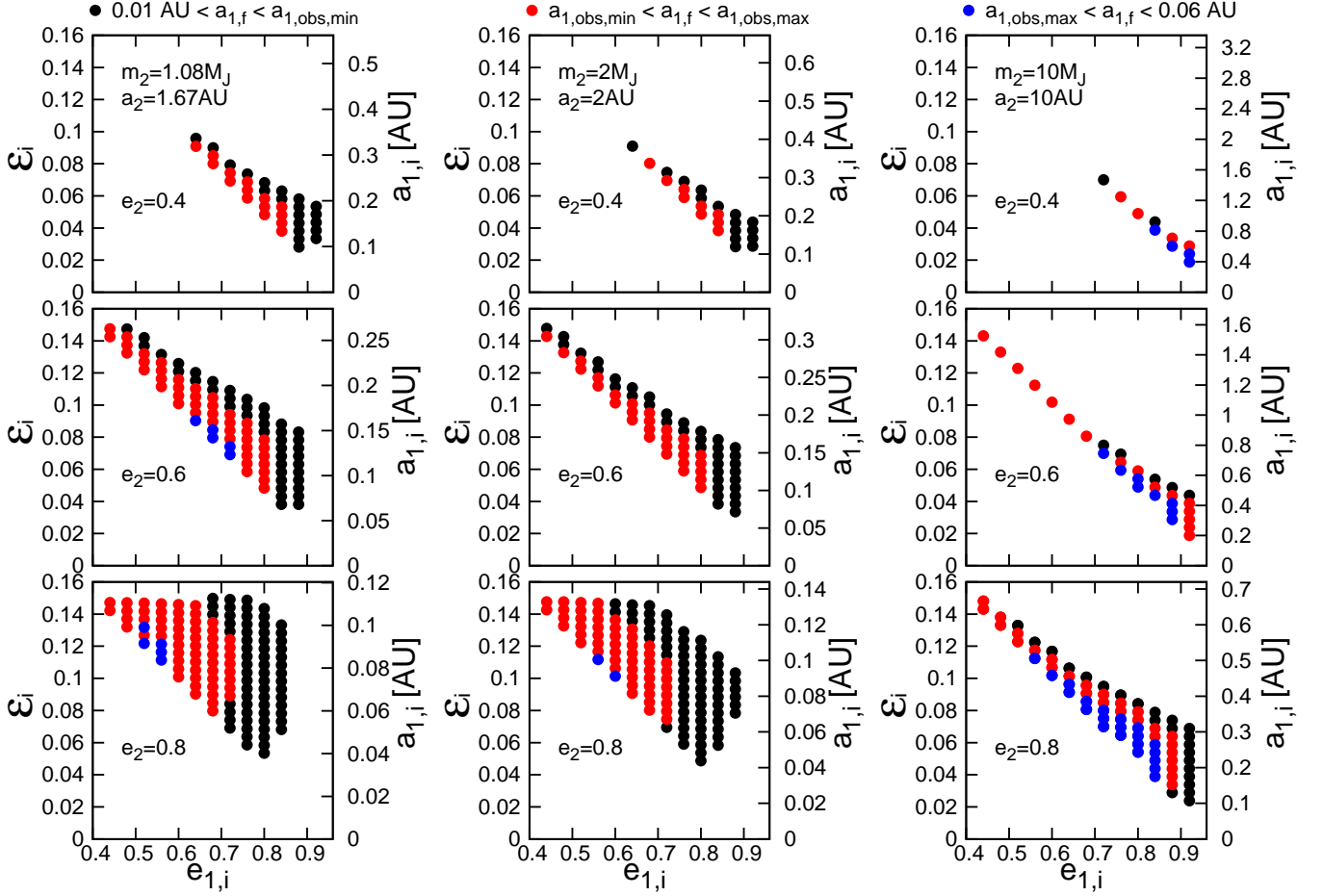} 
\caption{Resulting prograde super-Earths of Kepler-97 simulations on $(e_{1,i},\epsilon_i$) plane. Different colors correspond to different range of $a_{1,f}$ with black $0.01$ AU $<a_{1,f}<a_{1,\rm obs,min}$, red $a_{1,\rm obs,min}<a_{1,f}<a_{1,\rm obs,max}$, and blue $a_{1,\rm obs,max}<a_{1,f}<0.06$ AU, respectively, where $a_{1,\rm obs,min}$ and $a_{1,\rm obs,max}$ refer to the lower and upper limits in Table \ref{tab:data-obs}. All the models adopt $m_0=0.94M_{\odot}$, $m_1=3.5M_{\oplus}$, $i_{12,i}=6^{\circ}$, $i_{s1,i}=0^{\circ}$,  $t_{v,p}=0.001$ yr, and $f=2.44$, respectively. The initial values of $m_{2}$ and $a_{2}$ are $(m_{2},a_{2})=(1.08M_{\rm J}, 1.67$ AU), $(2M_{\rm J}, 2$ AU), and $(10M_{\rm J}, 10$ AU) from left to right. The initial value of $e_{2}$ is $0.4$, $0.6$, $0.8$, from top to bottom.} \label{fig:k97}
\end{center}
\end{figure}

Kepler-97 consists of a sun-like central star, a close-in super-Earth with $a\sim0.036$ AU likely in a circular orbit, and a recently detected distant outer perturber. The orbital elements of the outer perturber are not well determined and there are week constraints only; $m_2>344M_{\oplus}$ and $a_2>1.637$ AU \citep{Marcy2014}. Therefore, we consider three sets of parameters for $m_{2}$ and $a_{2}$, the results of which are plotted in Figure \ref{fig:k97}; $(m_{2},a_{2})=(1.08M_{\rm J}, 1.67$ AU), $(2M_{\rm J}, 2$ AU), and $(10M_{\rm J}, 10$ AU) from left to right. For each set of parameters, we run simulations with three different values of $e_{2,i}$; $e_{2,i}=0.4, 0.6,$ and $0.8$, corresponding to the top, middle, and bottom panels in Figure \ref{fig:k97}.

As before, we run the simulations for $\epsilon_i>\epsilon_{\rm L}$ estimated from equation (\ref{eq:epsilon-oct}). In this region, there are NMs, PSEs, and TDs.  In Figure \ref{fig:k97} we plot only PSEs with different colors according to $a_{1,f}$. All the results lead to $a_{1,f}\sim(0.01-0.06)$ AU, so we plot the simulations in three colors with black, red, and blue circles for the range of $0.01$ AU $<a_{1,f}<a_{1,\rm obs,min}$, $a_{1,\rm obs,min}<a_{1,f}<a_{1,\rm obs,max}$, and $a_{1,\rm obs,max}<a_{1,f}<0.06$ AU, respectively, where $a_{1,\rm obs,min}$ and $a_{1,\rm obs,max}$ refer to the lower and upper limits in Table \ref{tab:data-obs}.

As Figure \ref{fig:k97} indicates, the resulting PSE region becomes narrower as $m_2$ increases. Nevertheless, the range of $a_{1,f}$ is fairly insensitive to the choice of the orbital parameters. Independently of the orbital parameters, all the sets of simulations reproduce the observed range of $a_{1,f}$, {\it i.e.}, $0.036\pm0.007$ AU. Therefore we conclude that the current configuration of Kepler-97 can be explained over a wide range of parameters in CHEM.

\subsection{Other Hierarchical Triple Systems: Kepler-93, Kepler-407, HAT-P-2, HAT-P-4, HAT-P-17, and WASP-22 \label{sec:real-hitwo}}

\begin{figure}[t]
\begin{center}
\includegraphics[width=18cm]{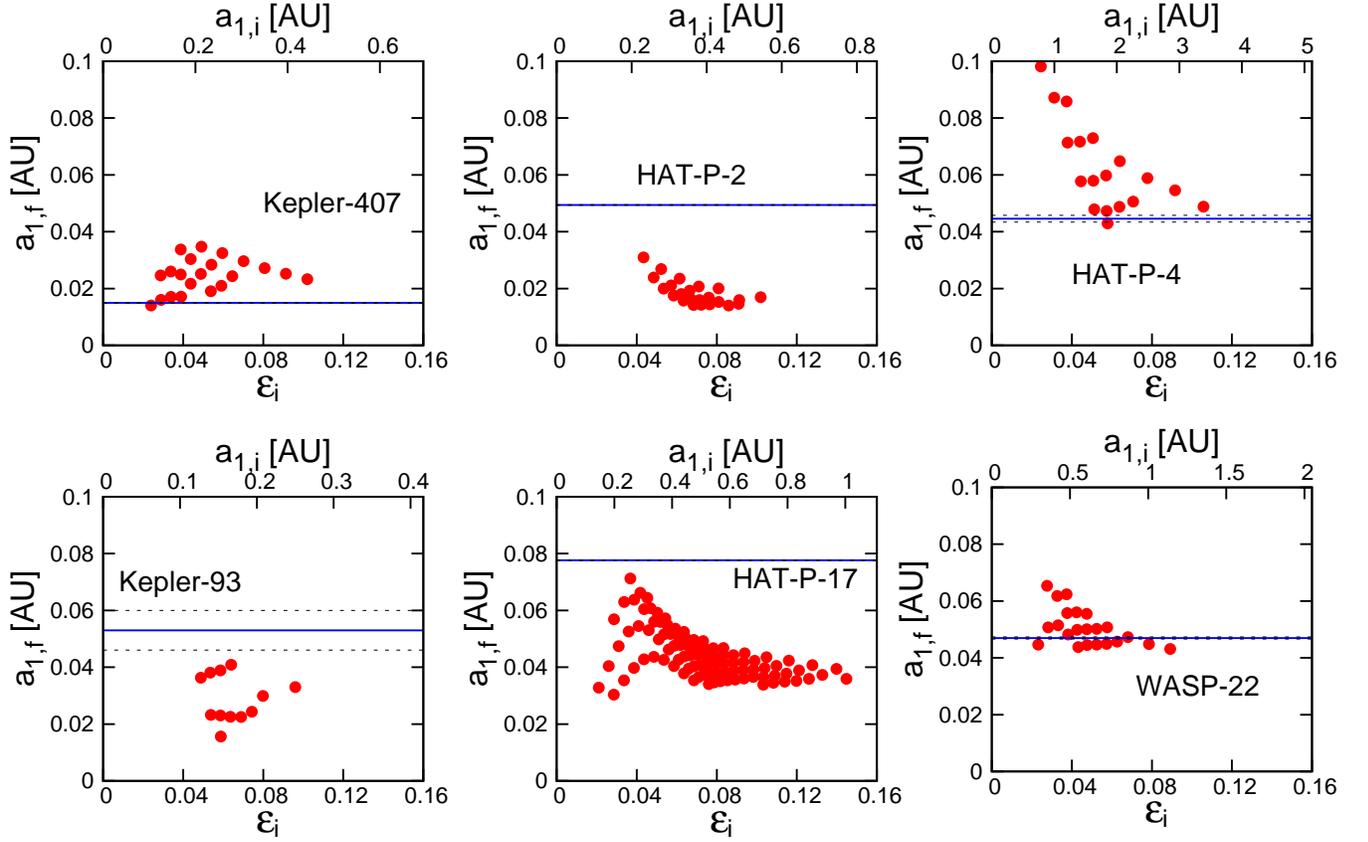} 
\caption{Final semi-major axis of resulting PHJs, $a_{1,f}$, for Kepler-93,  Kepler-407, HAT-P-2, HAT-P-4, HAT-P-17, and WASP-22 simulations against $\epsilon_i$. For each panel, the observed value, $a_{1,\rm obs}$, is plotted in the blue solid line and its lower and upper limits are shown in black dashed lines. All the models adopt $e_{2,i}=0.6$, $i_{12}=6^{\circ}$, $i_{s1}=0^{\circ}$, $t_{\rm v,p}=0.001$ yr, and $f=2.44$, respectively.} \label{fig:hjs}
\end{center}
\end{figure}

We repeat the same simulations on $(e_{1,i},\epsilon_{i})$ plane for the other six hierarchical triple systems and the results of $a_{1,f}$ against $\epsilon_i$ are plotted in Figure \ref{fig:hjs}. Each panel in Figure \ref{fig:hjs} exhibits several distinct sequences, which correspond to the different value of $e_{1,i}$; we plot the observed value of $a_{1.\rm obs}$ in blue solid lines and their lower and upper limits in black dashed lines. 

According to our simulations, three of them, Kepler-407, HAT-P-4, and WASP-22 are reproduced in CHEM, but the remaining other three systems, Kepler-93, HAT-P-2, and HAT-P-17 are not; the above three unsuccessful systems lead to $a_{1,f}$ smaller than their current values. As described in subsection \ref{sec:af-dis}, stronger tides are necessary for systems to achieve the larger $a_{1,f}$. The left bottom panel of Figure \ref{fig:af-dep} indicates that $a_{1,f}$ increases by $\sim30\%$ if $t_{\rm v,p}$ decreases by a factor of 10 from the fiducial value ($t_{\rm v,p}=0.03$ yr). In order to explain Kepler-93 and HAT-P-17 in the framework of CHEM, 10 times stronger tides are required. Such extreme tides seem to be unrealistic. Therefore, we conclude that those systems are unlikely to result from CHEM. 

We find that only 4 out of 7 are possibly reproduced, but the other 3 are difficult. Although the 7 systems we consider here may not represent the fair sample of the hierarchical triple systems, this may imply that CHEM can reproduce a reasonable fraction of close-in prograde planets in the observed hierarchical triple systems, but it is also likely that the other migration processes should operate including the Lidov-Kozai migration and disk migration.  

\subsection{Counter-orbiting HJ Systems: HAT-P-6 and HAT-P-14\label{sec:real-coHJ}}

In this subsection, we apply CHEM to two counter-orbiting HJ systems, HAT-P-6 and HAT-P-14. Since the possible outer perturbers for these systems are not confirmed, we assume a hypothetical outer perturber in a hierarchical triple configuration and examine if the counter-orbiting HJ can be reproduced in this scenario. Again, we repeat the same simulations on ($e_{1,i},\epsilon_{i}$) plane. The hypothetical perturber we adopt has $a_{2,i}=1000$ AU, $m_{2}=0.03M_{\odot}$, and $e_{2,i}=0.6$. Although there are a variety of possible configurations for the outer perturber, we follow a set of parameters for a sub-stellar perturber according to \citet{XueSuto2016}. \citet{XueSuto2016}, however, found that the formation of counter-orbiting HJs is difficult for such a configuration. In our simulations, we increase the formation efficiency by adopting less restrictive disruption radius; we use the disruption control factor $f=2.16$, instead of the fiducial value $f=2.7$.

\begin{figure}[t]
\begin{center}
\includegraphics[width=18cm]{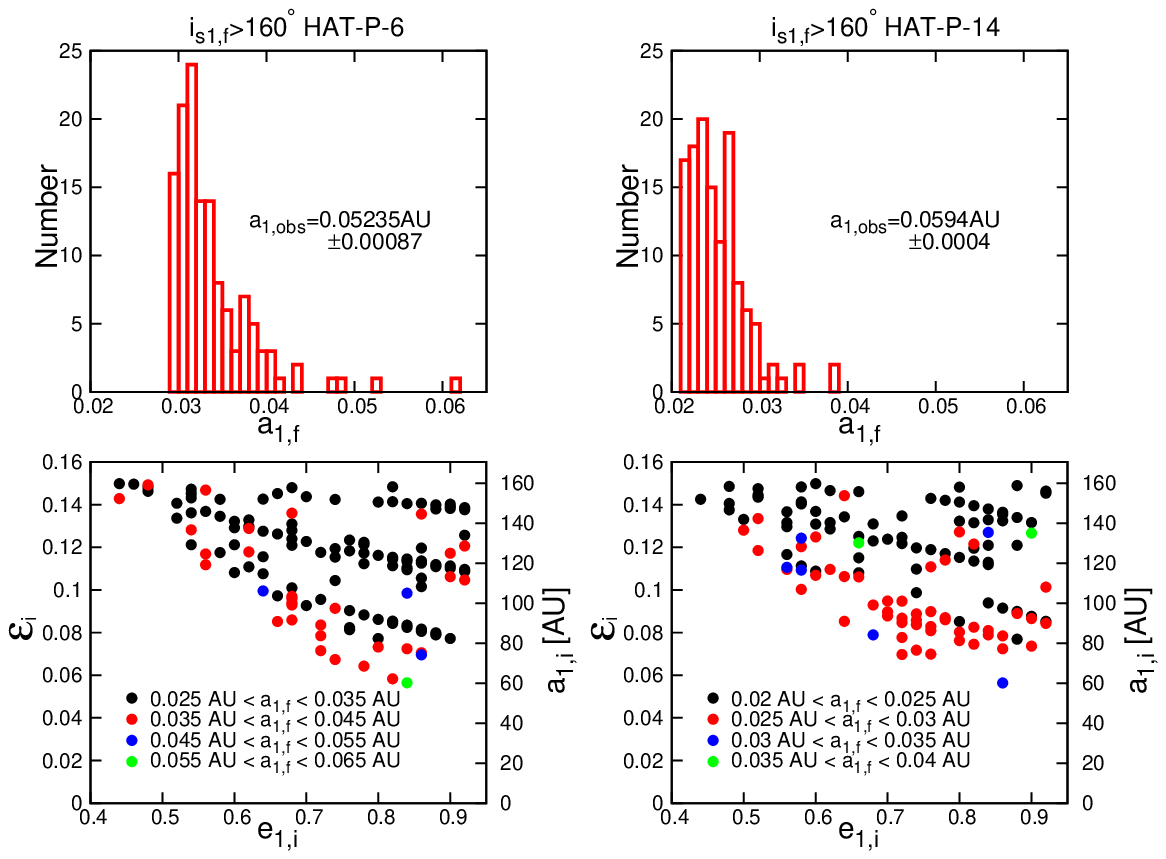} 
\caption{Distribution of $a_{1,f}$ for resulting RHJs with $i_{s1}>160^{\circ}$ in HAT-P-6 and HAT-P-14 simulations ({\it upper}) and those RHJs on ($e_{1,i},\epsilon_i$) plane ({\it bottom}). Different colors correspond to different $a_{1,f}$. All the models adopt $m_{2}=0.03M_{\odot}$, $a_{2}=1000$ AU, $e_{2,i}=0.6$, $i_{12,i}=6^{\circ}$, $i_{s1,i}=0^{\circ}$, $t_{\rm v,p}=0.03$ yr, and $f=2.16$. Left: HAT-P-6 with black $0.025<a_{1,f}<0.035$ AU, red $0.035<a_{1,f}<0.045$ AU, blue $0.045<a_{1,f}<0.055$ AU, and green $0.055<a_{1,f}<0.065$ AU; Right: HAT-P-14 with black $0.02<a_{1,f}<0.025$ AU, red $0.025<a_{1,f}<0.03$ AU, blue $0.03<a_{1,f}<0.035$ AU, and green $0.035<a_{1,f}<0.04$ AU.} \label{fig:count}
\end{center}
\end{figure}

The results are shown in Figure \ref{fig:count}. We plot the distribution of $a_{1,f}$ for RHJs with final spin-orbit angle $i_{s1,f}>160^{\circ}$ and those RHJs on ($e_{1,i},\epsilon_{i}$) plane in the upper and bottom panels for above two systems. We use different colors in the bottom panels according to $a_{1,f}$. The left panels refer to HAT-P-6. In this case, we adopt black for $0.025<a_{1,f}<0.035$ AU, red for $0.035<a_{1,f}<0.045$ AU, blue for $0.045<a_{1,f}<0.055$ AU, and green for $0.055<a_{1,f}<0.065$ AU. While a non-negligible fraction of counter-orbiting HJ systems are produced, only one case satisfies the current observation with $a_{1,i}=60.2$ AU, $e_{1,i}=0.84$, $a_{1,f}=0.0523$, and $i_{s1,f}=165.1^{\circ}$. Most of RHJs have $a_{1,f}$ smaller than $a_{1,\rm obs}$. In conclusion, it is difficult to produce counter-orbiting HJ systems that are consistent with the current observation of HAT-P-6.

The right panels are for HAT-P-14. In this case, we choose black for $0.02<a_{1,f}<0.025$ AU, red for $0.025<a_{1,f}<0.03$ AU, blue for $0.03<a_{1,f}<0.035$ AU, and green for $0.035<a_{1,f}<0.04$ AU. The maximum $a_{1,f}$ in the simulations, 0.038 AU, is only $\sim65\%$ of $a_{1,\rm obs}=0.0594$ AU. Therefore, it is completely impossible for HAT-P-14 to be reproduced in CHEM.

These results are supposed to be generic regardless of the orbital parameters of the outer perturber, because the range of $a_{1,f}$ is fairly insensitive to them, if we assume the tendency of $a_{1,f}$ in RHJs is similar as in PHJs shown in Figure \ref{fig:af-dep}. We expect that in general it is very difficult to form counter-orbiting HJs without fine tuning, so we conclude that even CHEM is difficult to explain the observed candidates of counter-orbiting HJ systems. This implies that the observed candidate counter-orbiting HJ systems may be simply due to the projection effect like HAT-P-7 or another physical mechanism other than CHEM is responsible to produce them. This is still an open question.

\begin{sidewaystable}[t]
\begin{center}
\caption{Orbital parameters of selected hierarchical triple and candidate counter-orbiting HJ systems.}
\begin{threeparttable}
\begin{tabular}{|c||c|c|c|c|c|c|c|c|} 
\hline
&$m_{0}$& $m_{1}$ & $m_{2}$&$a_{1}$& $a_{2}$&$e_{1}$&$\lambda$ &Ref.\\ 
&$M_{\odot}$&&&AU&AU&&&\\  \hline
Hierarchical&&&&&&&&\\
triple &&&&&&&& \\\hline
Kepler-97& $0.94\pm0.06$ &$3.5\pm1.9M_{\oplus}$ & $>344M_{\oplus}$&$0.036\pm0.007$&$>1.637$&0.0&\nodata&1\\ \hline
Kepler-93& $0.91\pm0.06$ &$2.59\pm2.0M_{\oplus}$ & $>954M_{\oplus}$&$0.053\pm0.007$&$>2.441$&0.0&\nodata&1\\ \hline
Kepler-407& $1.0\pm0.06$ &$<3.1999M_{\oplus}$ & $4000\pm2000M_{\oplus}$&$0.01497\pm0.0003$&$4.068_{-0.466}^{+0.441}$&0.0&\nodata&1\\ \hline
HAT-P-2& $1.34\pm0.09$ &$8.74\pm0.26M_{\rm J}$ & $8-200M_{\rm J}$&$0.0674\pm0.00081$&4-31&$0.517\pm0.0033$&$9^{\circ}\pm13.4^{\circ}$&2,3\\\hline
HAT-P-4& $1.26\pm0.14$ &$0.68\pm0.04M_{\rm J}$ & $1.5-310M_{\rm J}$&$0.0446\pm0.0012$&5-60&0.0&$-4.9^{\circ}\pm11.9^{\circ}$&2,4\\\hline
HAT-P-17& $0.857\pm0.039$ &$0.534\pm0.018M_{\rm J}$ & $2.8-3.7M_{\rm J}$&$0.0882\pm0.00147$&4.7-8.3&$0.346\pm0.007$&${19^{\circ}}^{+14^{\circ}}_{-16^{\circ}}$&2,5\\\hline
WASP-22& $1.1\pm0.3$ &$0.56\pm0.103M_{\rm J}$ & $7-500M_{\rm J}$&$0.04698\pm0.00037$&6-40&0.0&$22^{\circ}\pm16^{\circ}$&2,6\\\hline
Counter- &&&&&&&&\\  
orbiting HJs &&&&&&&& \\  \hline
HAT-P-6&$1.29\pm0.06$&$1.057\pm0.119M_{\rm J}$&\nodata&$0.05235\pm0.00087$&\nodata&0.0&$165^{\circ}\pm6^{\circ}$&7\\\hline
HAT-P-14&$1.386\pm0.045$&$2.2\pm0.04M_{\rm J}$&\nodata&$0.0594\pm0.0004$&\nodata&0.0&$189.1^{\circ}\pm5.1^{\circ}$&8\\\hline
\end{tabular}
\begin{tablenotes}[para,flushleft]
\item{\textbf{Note.} The observations reported at http://exoplanet.org include data from the references as follows: (1) \citet{Marcy2014}; (2) \citet{Knutson2014}; (3)\citet{Pal2010}; (4) \citet{Kovacs2007}; (5) \citet{Howard2012}; (6)\citet{Maxted2010}; (7) \citet{Noyes2008}; (8) \citet{Torres2010}.} 
\end{tablenotes}
\end{threeparttable}
\label{tab:data-obs}
\end{center}
\end{sidewaystable}

\begin{table}[t]
\begin{center}
\caption{Orbital parameters of hierarchal triple and counter-orbiting HJ systems in our simulations.}
\begin{threeparttable}
\begin{tabular}{|c||c|c|c|c|c|c|} 
 \hline
&$m_{0}$& $m_{1}$ & $m_{2}$& $a_{2}$&$e_2$&Y/N \\ 
 &$M_{\odot}$&&&AU&& \\  \hline
Hierarchical&&&&&&\\
triple &&&&&&\\ \hline
Kepler-97v1& $0.94M_{\odot}$ &$3.5M_{\oplus}$ & $344M_{\oplus}$&$1.637$&0.4,0.6,0.8&$\surd$\\ \hline
Kepler-97v2& $0.94M_{\odot}$ &$3.5M_{\oplus}$ & $2M_{\rm J}$&$2$&0.4,0.6,0.8&$\surd$\\ \hline
Kepler-97v3& $0.94M_{\odot}$ &$3.5M_{\oplus}$ & $10M_{\rm J}$&$10$&0.4,0.6,0.8&$\surd$\\ \hline
Kepler-93& $0.91M_{\odot}$ &$2.59M_{\oplus}$ & $954M_{\oplus}$&$2.441$&0.6&$\times$\\ \hline
Kepler-407& $1.0M_{\odot}$ &$1.0M_{\oplus}$ & $4000M_{\oplus}$&4.068&0.6&$\surd$\\ \hline
HAT-P-2& $1.34M_{\odot}$ &$8.74M_{\rm J}$ & $60M_{\rm J}$&5&0.6&$\times$\\\hline
HAT-P-4& $1.26M_{\odot}$ &$0.68M_{\rm J}$ & $100M_{\rm J}$&30&0.6&$\surd$\\\hline
HAT-P-17& $0.857M_{\odot}$ &$0.534M_{\rm J}$ & $3.3M_{\rm J}$&6.5&0.6&$\times$\\\hline
WASP-22& $1.1M_{\odot}$ &$0.56M_{\rm J}$ & $30M_{\rm J}$&12&0.6&$\surd$\\\hline
Counter-&&  & & & &\\ 
orbiting HJs&&  & & &&\\ \hline
HAT-P-6&$1.29M_{\odot}$&$1.057M_{\rm J}$&$0.03M_{\odot}$&1000&0.6&$\surd$\\\hline
HAT-P-14&$1.386M_{\odot}$&$2.2M_{\rm J}$&$0.03M_{\odot}$&1000&0.6&$\times$\\\hline
\end{tabular}
\begin{tablenotes}[para,flushleft]
\item{\textbf{Note.} $m_0$, $m_1$, $m_2$, and $a_2$ are fixed to be consistent with Table \ref{tab:data-obs}. $e_2$ is basically set to $0.6$, and varied to $0.4$ and $0.8$ in Kepler-97 simulations to see the parameter dependence. The column Y/N indicates if the observed range can be recovered by the simulations or not with symbols $\surd$ and $\times$.} 
\end{tablenotes}
\end{threeparttable}
\label{tab:data-sim}
\end{center}
\end{table}

\section{Summary and Discussion \label{sec:discussion}}

Several formation and migration mechanisms have been proposed to explain the close orbit of HJs. None of them is fully consistent with observations, but it is also likely that each  of them has contributed to the observed HJs to some degree. In this paper, we have investigated a possibility that HJs form in near-coplanar eccentric hierarchical triple systems via the secular interaction between an inner planet and an outer perturber (CHEM). Our results are summarized in the following 5 main findings.

1) We generalize the analytic extreme eccentricity condition in purely gravitational interaction that was derived by \citet{Li2014} and \citet{Petrovich2015b} neglecting the mass of the inner planet, $m_1$, on the dynamics of the central star, $m_0$. We find that the extreme eccentricity region is significantly limited when the finite mass of $m_1$ is taken into account. Therefore, the significant migration in CHEM is possible only when $m_1\ll m_0$ and $m_1\ll m_2$.

2) We perform a series of numerical simulations in CHEM for systems consisting of a sun-like central star, giant gas inner planet, and planetary outer perturber, including the short-range forces (GR, stellar and planetary non-dissipative tides, and stellar and planetary rotational distortion) and stellar and planetary dissipative tides. We find that most of such systems experience a tidal disruption of the inner planet; a small fraction of prograde HJs are produced, but no retrograde HJ forms (see Figure \ref{fig:plfid}). The short-range forces suppress the extreme eccentricity growth, and significantly affect the orbital evolution of the inner planet. These results are fairly independent of the orbital elements of the outer perturber. 

3) We present an analytical model that explains the numerical results approximately. We analytically estimate the location when the minimum pericenter distance of the inner planet, $q_{1,\rm min}$, reaches the Roche limit, which determines the condition of the inner planet being tidally disrupted. In addition, we provide an analytical estimate of the migration time scale including short-range forces and planetary dissipative tides, which qualitatively explains the result of our numerical simulation. These estimates are useful in interpreting the simulation results.

4) We apply CHEM to super-Earth systems around a sun-like central star with a giant gas planetary outer perturber. As in the giant gas inner planetary case, we find that the majority end up with tidally disrupted planets, but a small fraction is survived as a prograde close-in super-Earth.

5) We apply CHEM to the observed 7 hierarchical triple systems (Kepler-93, Kepler-97, Kepler-407, HAT-P-2, HAT-P-4, HAT-P-17, and WASP-22) and 2 counter-orbiting HJ systems (HAT-P-6 and HAT-P-14). We find that 4 out of 7 hierarchical triple systems (Kepler-93, Kepler-407, HAT-P-4, and WASP-22) are possibly reproduced in CHEM, but the other 3 hierarchical triple systems and 2 counter-orbiting HJ systems are unlikely to be explained in CHEM.


Unfortunately, our simulations cannot predict the real fraction of HJs, because our parameter survey is performed in a biased fashion. First, we limit the parameter space of our simulations to the extreme eccentricity region, since the systems outside the extreme eccentricity region do not exhibit significant orbital migration. Second, we restrict our simulations in the range where the secular approximation in octupole expansion is valid (i.e., $\epsilon_i<0.15$ as described in section \ref{sec:initial}). Finally, our parameter survey is not realistic because we do not consider the prior distribution for the orbital parameters of the inner planets and outer perturber. Nevertheless, our main conclusion basically holds: CHEM can produce some fraction of close-in prograde planets, but no retrograde one. 

Our current conclusion seems to be slightly different from the case of a sub-stellar perturber. Indeed, \citet{XueSuto2016} found that very small fraction may end up with retrograde HJs. This is simply because of the initial semi-major axis of the inner planet, $a_{1,i}$. The range of $a_{1,i}$ is very different in the two models; we consider $0.3<a_{1,i}<8$ AU, while \citet{XueSuto2016} considered $3<a_{1,i}<80$ AU. In the former case, the inner planet suffers from tidal circularization before acquiring the required extreme eccentricity for the orbital flip, and therefore, no retrograde HJ forms in a planetary perturber.

Finally, the range of the final semi-major axis of HJs, $a_{1,f}$, may be a potential discriminator for their formation mechanisms. Our simulation indicates the range of $a_{1,f}$ for HJs is mainly determined by the strength of tides. Since in any high-$e$ migration mechanism, $a_{1,f}$ should be similarly determined by the competition between the eccentricity growth and the strength of tides, we expect that the resulting range of $a_{1,f}$ is common for any high-$e$ migration mechanism. On the other hand, disk migration would predict a very different $a_{1,f}$ because tides are not so important. Therefore, the range of $a_{1,f}$ can potentially distinguish between high-$e$ migration and disk migration. In section \ref{sec:real}, we find all the three unsuccessful hierarchical triple systems, (Kepler-93, HAT-P-2, and HAT-P-17) have the semi-major axis of the inner planet smaller than their observed values. This implies that those three systems are difficult to form by any high-$e$ migration mechanism, unless tides are unrealistically efficient. Alternatively, those systems may result from disk migration. It is also tempting to apply this prediction of the range of $a_{1,f}$ for other observed close-in planetary systems with a hierarchical triple configuration, so as to estimate the possibility of those systems resulting from high-$e$ migration.

\acknowledgments

We thank Gongjie Li for useful discussions and an anonymous referee for various constructive suggestions. K.M. is supported by the Leading Graduate Course for Frontiers of Mathematical Sciences and Physics (FMSP).  This work is supported by JSPS Grant-in-Aids for Scientific Research No. 26-7182 to K.M.,  and No. 24340035 to Y.S.  This work is supported in part by JSPS Core-to-Core Program ``International Network of Planetary Sciences". This work was performed in part under contract with the California Institute of Technology (Caltech)/Jet Propulsion Laboratory (JPL) funded by NASA through the Sagan Fellowship Program executed by the NASA Exoplanet Science Institute.



\end{document}